\documentclass[aps,prb,amsmath,amssymb,floatfix,twocolumn]{revtex4}
\usepackage{bbold}
\usepackage{graphicx}
\usepackage{subfigure}
\usepackage{color}
\usepackage{hyperref}

\begin{document}

\title{Topological properties of possible Weyl superconducting states of URu$_\mathbf{2}$Si$_\mathbf{2}$}

\author{Pallab Goswami}
\affiliation{National High Magnetic Field Laboratory and Florida State University, Tallahassee, Florida 32310, USA}

\author{Luis Balicas}
\affiliation{National High Magnetic Field Laboratory and Florida State University, Tallahassee, Florida 32310, USA}

\begin{abstract}
We show that the current thermodynamic measurements in the superconducting phase of $\mathrm{U}\mathrm{Ru}_2\mathrm{Si}_2$ are compatible with two distinct singlet chiral paired states $k_z(k_x \pm i k_y)$ and $(k_x \pm i k_y)^2$. Depending on the locations of the Fermi surface in the hidden ordered phase, both of these chiral d-wave pairings can support point and line nodes. Despite possessing similar low temperature thermodynamic properties, these two pairings are topologically distinguished by their respective orbital angular momentum projections along the c-axis, $m=\pm 1$ and $m=\pm 2$. The point nodes of these states act as the monopoles and the anti-monopoles of the Berry's gauge flux of charge $\pm m$, which are separated in the momentum space along the $c$ axis. As a result, the Berry's flux through the $ab$ plane equals $m$. Consequently, the point nodes of $k_z(k_x+i k_y)$ and $(k_x \pm ik_y)^2$ states respectively realize the Weyl and the double-Weyl fermions, with chemical potential exactly tuned at the Fermi point, due to the charge conjugation symmetry. These topologically nontrivial point nodes, give rise to $m$ copies of protected spin degenerate, chirally dispersing surface states on the $ca$ and the $cb$ planes, which carry surface current, and their energies vanish at the Fermi arcs. In contrast, a line node acts as the momentum space vortex loop, and gives rise to the zero energy, dispersionless Andreev bound states on the surfaces parallel to the plane enclosed by the line node. The Berry's flux through the $ab$ plane gives rise to anomalous spin Hall and thermal Hall conductivities, and various magnetoelectric effects. A clear determination of the bulk invariant can only be achieved by probing the pairing symmetry via a corner Josephson junction measurement, and Fourier transformed STM measurements of the Fermi arcs. Therefore, we identify $\mathrm{U}\mathrm{Ru}_2\mathrm{Si}_2$ as a promising material for realizing gapless topological superconductivity in three spatial dimensions.
\end{abstract}

\maketitle

\section{Introduction}\label{sec1}
The neutrinos have been originally thought to be chiral massless particles, and within the framework of the Standard Model these elementary, charge neutral particles have been traditionally described by the left handed Weyl fermions. However, the very notion of the massless neutrinos has now become defunct due to the observation of neutrino oscillations. Therefore, a fundamental question still remains, if the Weyl fermion is indeed realized in nature. In this regard, condensed matter systems serve as the most promising ground, where the massless, chiral fermions may be realized as the low energy quasi-particles in diverse systems\cite{Volovik1,Vishwanath,Xu,Burkov1,Burkov2,Zyuzin1,Cho,Meng,Gong,Sau,Das}. Unlike in high energy physics, the Weyl fermions in condensed matter systems can be charged as well as charge neutral, and a variety of anomalous chiral properties can be probed with table-top experiments\cite{Ninomiya,Aji,SonSpivak,Kharzeev,Landsteiner,Ran,Zyuzin3,Grushin,Qi,Goswami2,Burkov3,Goswami3,Chernodub}.

In the simplest form, a pair of Weyl fermions are described by the following Hamiltonians in the momentum space $ H_{\pm}= \pm \hbar v \; \boldsymbol \sigma \cdot (\mathbf{k}-\mathbf{k}_{\pm})$, where $\pm$ respectively correspond to the chirality of the right and the left-handed fermions, and $\mathbf{k}_{\pm}$ denote the band-touching points. We have denoted the three Pauli matrices by $\boldsymbol \sigma$, which can operate on the physical or the pseudo-spin, depending on the physical context. Due to the gapless nature, the Weyl semi-metal describes a quantum critical system with dynamic exponent $z=1$, even though the quasiparticle residue is finite \cite{Goswami1}. In the momentum space, the band-touching points for the right and the left handed fermions respectively have the singular forms of a hedgehog and an anti-hedgehog, of topological charges $q=\pm 1$. Owing to these momentum space singularities, the chiral fermions appear as the sink and the source of the Berry's gauge flux. This is how the Weyl semi-metal marries the disjoint notions of the fermionic quantum criticality and the momentum space topology \cite{Goswami3}. As a result of the topological non-triviality, the Weyl fermions support protected, chirally dispersing surface states, on the particular surfaces whose normals are orthogonal to the separation of the Weyl fermions. The dispersions of the surface states vanish along a line in the surface Brillouin zone, which is bounded by the images of the monopole and the anti-monopole, and is known as the Fermi arc. When the nodal separation vanishes, we obtain a four component massless Dirac fermion, which is a topologically trivial quantum critical system, and concomitantly the surface states disappear. In condensed matter physics, one can also find chiral quasiparticles, which are described by the (anti)hedgehog with the integer topological charge $q > 1$, and there is no analog of such quasiparticles in the realm of Lorentz invariant description of high energy physics.

When there is only one pair of Weyl fermions, and the momentum space separation of the nodal points $\Delta \mathbf{K} = \mathbf{k}_+-\mathbf{k}_-$ is non-zero, the system violates the time-reversal symmetry. The low energy Hamiltonian acquires the form
\begin{equation}
\hat{H}_{W}=\begin{bmatrix} \hbar v\boldsymbol \sigma .\left(\mathbf{k}+\frac{\delta\mathbf{K}}{2}\right) & 0  \\ 0  & - \hbar v\boldsymbol \sigma .\left(\mathbf{k}-\frac{\delta\mathbf{K}}{2}\right) \end{bmatrix},
\label{eq:1}
\end{equation}
and the $\Delta \mathbf{K}$ appears as a constant, axial vector potential. This separation determines the anomalous Hall responses of the appropriate conserved quantity, which can be understood in terms of Berry flux through the plane perpendicular to $\Delta \mathbf{K}$  or the axial anomaly of the Weyl fermions \cite{Ran,Zyuzin3,Grushin,Qi,Goswami2,Goswami3}. The anomalous Hall conductivity of the conserved charge $g$ is given by
\begin{equation}\label{anomaly}
\sigma_{ab,g}=\epsilon_{abc} \; \frac{g^2}{2 \pi h} \Delta K_c.
  \end{equation}In the absence of superconductivity, the electric charge $g=e$ is conserved and we expect anomalous charge Hall effect. For a superconductor the electric charge is not a conserved quantity due to the violation of the electromagnetic gauge symmetry. But, the energy is a conserved quantity and we expect an anomalous thermal Hall effect. Depending on whether the quasiparticles are complex Weyl fermions or real Majorana-Weyl fermions, the thermal Hall conductivities at low temperatures are respectively obtained by setting $g^2=\pi^2 k^2_B T/3$ and $g^2=\pi^2 k^2_B T/6$ (see Ref.~\onlinecite{Goswami2}). If the spin is also a conserved quantity, as in the case of a singlet superconductor, the anomalous spin Hall conductivity is found by setting $g=\hbar/2$. There can be additional chiral transport phenomena, if the left and the right handed Weyl points are separated in energy \cite{Kharzeev,Landsteiner,Zyuzin3,Grushin,Qi,Goswami2,Goswami3,Chernodub}. Therefore, the chiral quasiparticles in condensed matter system can serve as a platform for realizing field theory anomalies, and also for exploring the exotic anomalous transport properties for future technological applications.

For this reason there are considerable ongoing theoretical and experimental efforts in identifying three dimensional chiral fermions in real materials. It is becoming increasingly clear that the strong spin orbit coupling combined with correlation effects can produce such exotic states \cite{Vishwanath,Krempa}. In this respect, the heavy fermion compounds are very promising materials, due to naturally present strong spin orbit coupling and electronic correlations. The correlation effects usually give rise to various broken symmetry states, such as spin-density wave and unconventional superconductivity. In this paper we show that depending on the actual pairing symmetry, the superconducting state of URu$_2$Si$_2$, can realize either the Weyl fermions or the double-Weyl fermions, which are respectively the unit or the double Berry (anti)monopoles.
\begin{figure}[htb]
\includegraphics[width=8.5cm,height=6.5cm]{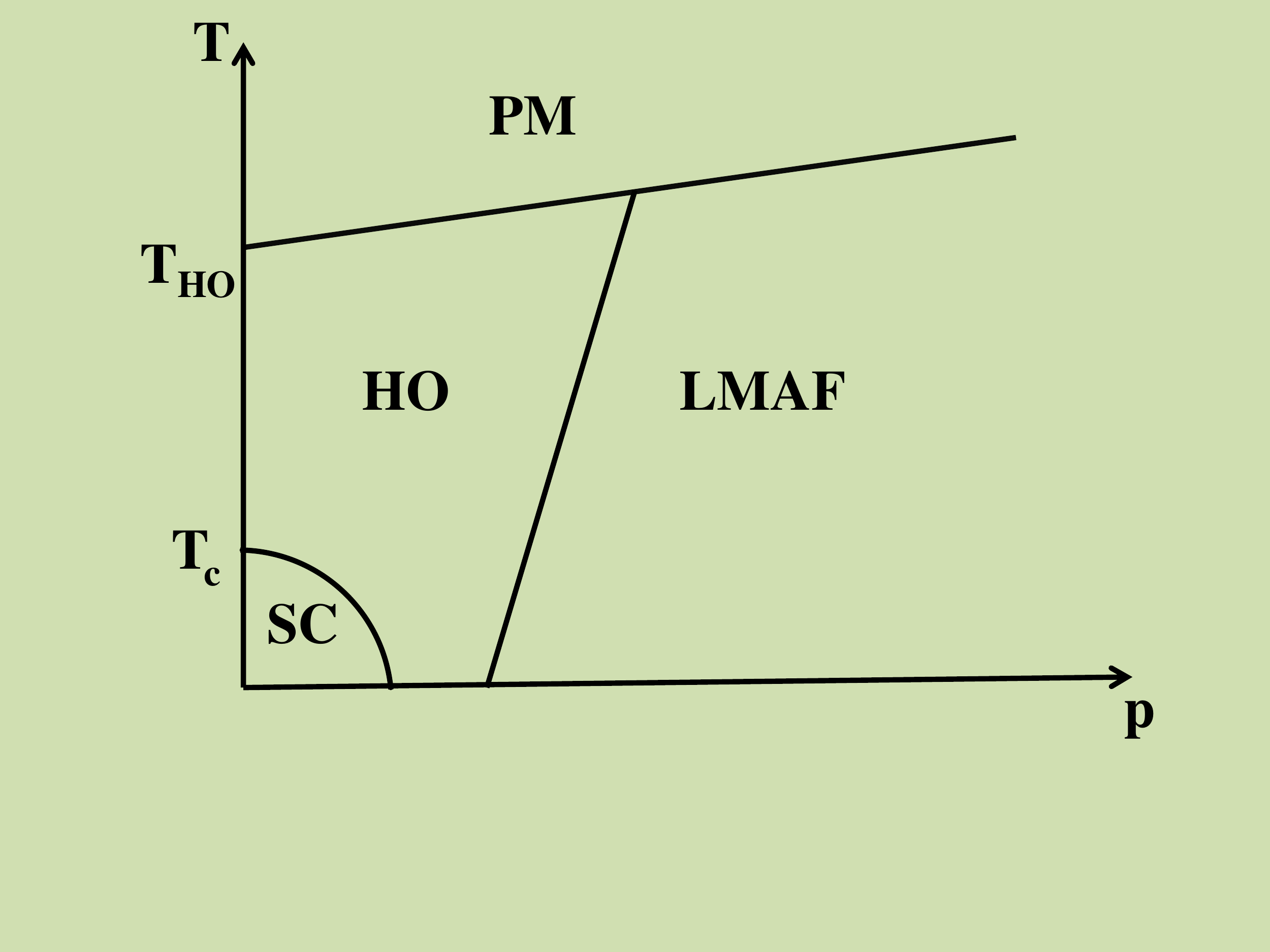}
\caption{(Color online)The phase diagram of URu$_2$Si$_2$ as a function of temperature T and applied pressure p. The four phases disordered paramagnetic metal, hidden ordered state, large moment antiferromagnet and superconductor are respectively denoted by PM, HO, LMAF, and SC. The transition temperatures are $T_{HO}\sim 17.5$ K and $T_c \sim 1.5$K.}
\label{fig:1}
\end{figure}

\subsection{Possible pairing symmetries for URu$_2$Si$_2$}
URu$_2$Si$_2$ is a remarkable strongly correlated material, and its phase diagram in the temperature-pressure plane consists of four phases, as depicted in  Fig.~\ref{fig:1}. At high temperatures we have a paramagnetic, metallic phase (PM), without any broken symmetry. At ambient pressure and at $T_{HO}=17.5$ K, URu$_2$Si$_2$ undergoes a second order transition into an enigmatic hidden ordered (HO) phase, which can be destabilized in favor of a large moment antiferromagnetic (LMAF) phase beyond a threshold value of the applied pressure. In addition there is an unconventional superconducting state (SC) at very low temperatures \cite{Palstra,Maple,Bredl}. At ambient pressure, the superconducting transition temperature $T_c \sim 1.5$ K, and it can be further lowered by the applying pressure. At present it is not entirely clear, if there is a direct transition between the SC and the LMAF states, or the SC state remains completely nestled inside the hidden ordered phase.
\begin{figure*}[htb]
\includegraphics[width=13cm,height=8cm]{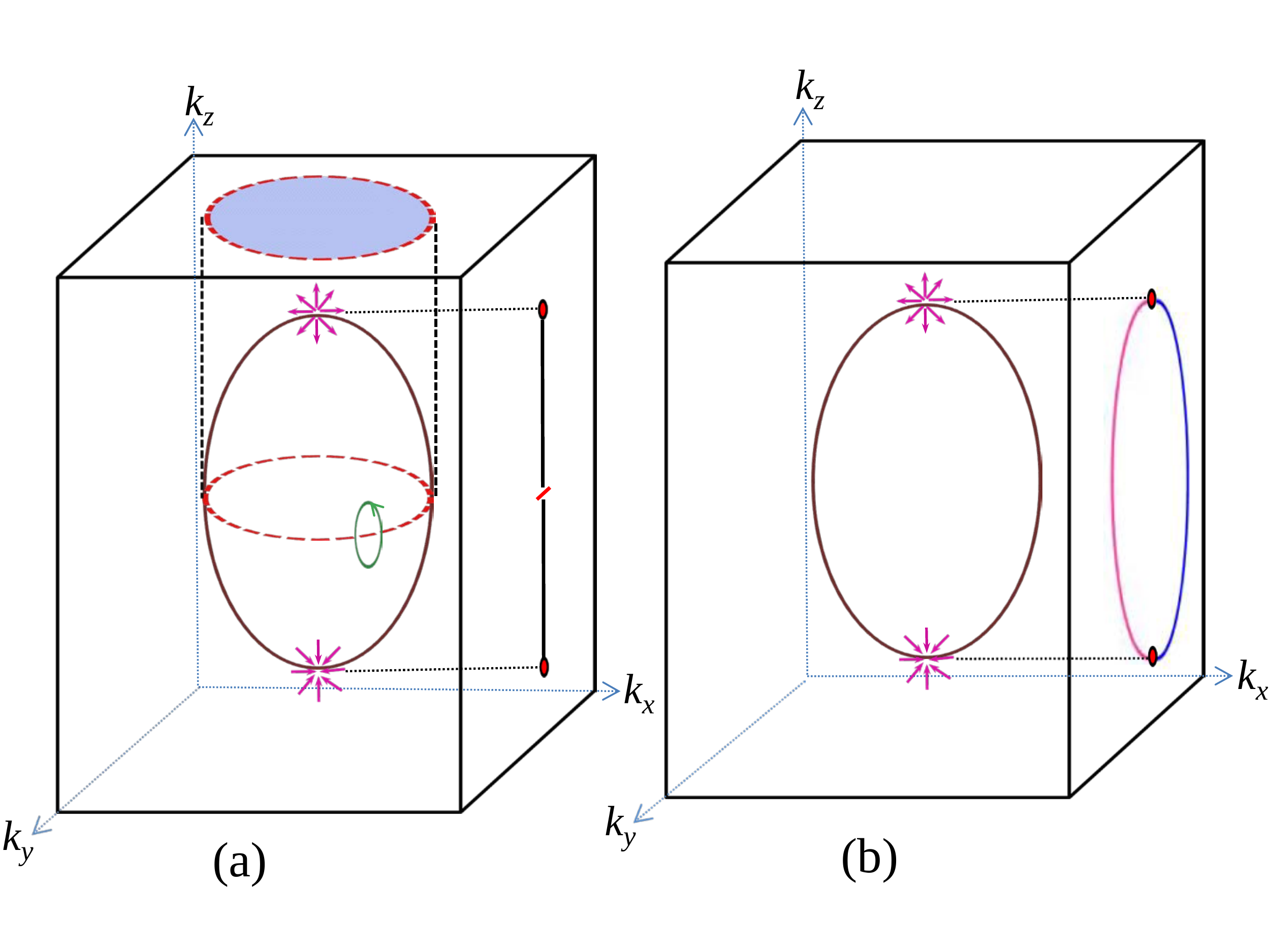}
\caption{(Color online) (a) The line and the point nodes of $k_z(k_x \pm i k_y)$ pairing. The wavefunction for the BCS quasiparticle changes sign while going around the line node along a closed path like the green curve with arrow. The zero energy, dispersionless Andreev bound states on the $(0,0,1)$ surface due to the line node produce an image of the equatorial cross-section of the Fermi surface and is marked by the blue filled region. The point nodes act as unit charge Berry (anti)monopoles which give rise to the chiral surface states on the $(1,0,0)$ surface, and their energies vanish along the Fermi arc $k_y=0$, bounded by the images of the point nodes. There is no chiral mode for $k_z=0$, and this discontinuity of the Fermi arc is marked in red. (b) The bulk point nodes which act as double Berry (anti)monopoles and two independent Fermi arcs $k_y=\pm \sqrt{k^2_F-k^2_z}$ for $(k_x \pm i k_y)^2$ pairing on the $(1,0,0)$ surface. }
\label{fig:2}
\end{figure*}
The experimental and theoretical investigations have been primarily focused on unraveling the nature of the hidden ordered phase. But, in this paper we do not discuss the nature of the hidden ordered state and various theoretical proposals put forward towards explaining its puzzling phenomenology. A detailed survey of this topic can be found in Ref.~\onlinecite{Mydosh}. Despite an intense effort for more than two decades, there is no consensus on the actual nature of the hidden order parameter. In contrast, the superconducting state has received relatively little attention, and the experimental effort has remained sporadic. Inside the SC state, the low temperature specific heat $C_v$ and the nuclear magnetic relaxation rate $T^{-1}_1$ respectively show power law behaviors according to $C_v \propto T^2$ (see Ref.~\onlinecite{Hasselbach}) and $T^{-1}_1 \propto T^3$  (see Ref.~\onlinecite{Kohori}). These power law behaviors imply that the density of states for the nodal quasiparticles vanishes linearly with energy i.e., $D(\epsilon) \propto \epsilon$. The longitudinal thermal conductivity $\kappa_{xx}$ at zero and weak magnetic fields behave as, $\propto T$ and $\propto \sqrt{H}$ (see Ref.\onlinecite{Kasahara1,Kasahara2}) respectively. All of these power laws are easy to reconcile with the existence of line nodes. However, the thermal conductivity in the presence of the strong magnetic field (above $H_{c1}$) behave very differently depending on the direction of the applied field. For a field along the c-axis, after initial increase as $\sqrt{H}$, $\kappa$ saturates beyond some field strength $H_s$, and finally shows sharp decrease around the Pauli limited $H_{c2}$. In contrast, for an applied field along the $a$ axis, $\kappa_{xx}$ continues to grow beyond $H_s$, dropping sharply around $H_{c2}$. This has been construed as an  evidence for point nodes along the $c$ axis. These observations have motivated Kasahara {\it et al.} to propose a chiral d-wave paired state of the form
\begin{equation}\label{pairing1}
\Delta(\mathbf{k})=\Delta_{0}\sin \frac{c k_z}{2}\left[\sin a \frac{k_x}{2} \pm i \; \sin a \frac{k_y}{2}\right],
\end{equation}
which can give rise to the line nodes in the $ab$ plane, if the underlying Fermi pocket is located around the $\mathbf{k}=(0,0,0)$ ($\Gamma$) or the $\mathbf{k}=(0,0,2\pi/c)$ ($Z$) points\cite{Kasahara1,Kasahara2}. In addition there are point nodes at the poles of the Fermi pockets along the $c$ axis, which are linearly dispersing Weyl fermions with density of states $\propto \epsilon^2$. Conventional thermodynamic response and dissipative transport properties will be dominated by the line node due to higher density of states. The presence of point nodes have also been supported by the subsequent field angle dependent specific heat measurements \cite{Yano}. In addition, neutron scattering resonances have been observed inside the SC state, at a commensurate wave-vector $\mathbf{Q}=(0,0,2\pi/c)$ and also at some additional incommensurate wave-vectors \cite{Hassinger}. The neutron scattering resonance at the commensurate wave-vector suggests that $\Delta(\mathbf{k}+\mathbf{Q})=-\Delta(\mathbf{k})$. The proposed gap function in Eq.~(\ref{pairing1}) indeed satisfies this condition.

But, the most interesting aspect of this gap function is the violation of the time-reversal symmetry and the reflection symmetries about the $ca$, the $cb$, and the $ab$ planes. In the magnetization measurements a paramagnetic response has been found at high magnetic fields, which disappears at $H_{c2}$\cite{GangLi}. The observation of paramagnetic response inside the superconducting state is consistent with the presence of an underlying orbital magnetic moment originating from a time reversal symmetry breaking paired state. Very recently, a polar Kerr effect has been observed in the superconducting state, which shows the existence of a time reversal symmetry breaking, chiral paired state\cite{Kapitulnik}.

However, there are considerable uncertainties regarding the fermiology inside the HO state, and there can be other paired states, which have similar thermodynamic properties as $\Delta(\mathbf{k})$ in Eq.~(\ref{pairing1}). In particular we will show there is another contending chiral d-wave paired state
\begin{eqnarray}\label{pairing2}
\Delta^{\prime}(\mathbf{k})=\Delta_{0}\cos \frac{c k_z}{2}\bigg[\cos a \frac{k_x}{2} - \cos a \frac{k_y}{2}\pm i \; \sin a \frac{k_x}{2} \nonumber \\ \times \; \sin a \frac{k_y}{2}\bigg],
\end{eqnarray}
which possesses similar thermodynamic properties and the neutron scattering resonance at $\mathbf{Q}$. If there are Fermi pockets around $(0, 0, \pm \pi/c)$ this gap function can have line nodes. In addition there are point nodes along the $c$ axis. Interestingly, both the line and the point nodes of this gap function have linear density of states ($\propto \epsilon$) (see Sec.~\ref{sec3}). Therefore, it is almost impossible to unambiguously distinguish the above two pairings, based on the conventional thermodynamic or transport measurements. However, we note that the field angle dependent thermal conductivity, and the specific heat measurements have been claimed to support $\Delta(\mathbf{k})$ in Eq.~(\ref{pairing1}). A different chiral d-wave state has also been proposed in Ref.~\onlinecite{Hsu}, where only one component of the complex pairing amplitudes changes sign under $\mathbf{k} \to \mathbf{k} + \mathbf{Q}$, and gives rise to point nodes.

\subsection{Topological aspects of possible chiral pairings}
When we focus on the fermi pockets around the $\Gamma$ or the $Z$ points, the above pairings can be respectively approximated by
\begin{eqnarray}
\Delta(\mathbf{k}) \approx \frac{\Delta_{0}}{k^2_F} k_z (k_x \pm i \; k_y) \propto Y^{m=\pm1}_{l=2}(\hat{\Omega}), \\
\Delta^{\prime}(\mathbf{k}) \approx \frac{\Delta_{0}}{k^2_F} (k_x \pm i k_y)^2 \propto Y^{m=\pm2}_{l=2}(\hat{\Omega}),
\end{eqnarray}
where $Y^m_l(\hat{\Omega})$ are the spherical harmonics, and the $\hat{\Omega}$ is the solid angle defined on the Fermi pocket. Therefore, these two pairings are distinguished by their orbital angular momentum projections ($m$) along the $c$-axis. In the following sections, we demonstrate that all the topological aspects of the chiral pairing and the point nodes are entirely determined by $m$. In particular we show that the point nodes of the two pairings in Eq.~\ref{pairing1} and Eq.~(\ref{pairing2}) respectively act as the unit and the doouble Berry (anti)monopoles, and realize the Weyl and the double-Weyl fermions. The angular momentum $m$ of the Cooper pairs also determines the number of chiral surface states and Fermi arcs on the $ca$ and the $cb$ planes, as shown in Fig. 2. We find that the nontrivial Berry flux through the $ab$ plane equals $2 \pi m$, and is responsible for anomalous spin Hall, thermal Hall (as described by Eq.~(\ref{anomaly})), spin-Nernst effects, and magnetoelectric effects. In particular we show that a large anomalous thermal Hall conductivity (the maximum value of the thermal Hall conductivity) $ \sim 10^{-3} WK^{-1}m^{-1}$ is achieved around $T \sim T_c/2$.

We also show that the line node of the gap function acts as a vortex loop in the momentum space, and possesses non-trivial momentum space topology. However, the topological invariant of the line node does not depend on the angular momentum $m$, and is independent of the chirality. The line node of $k_z(k_x \pm i k_y)$ pairing leads to dispersionless, zero energy Andreev bound states on the $(0,0,1)$ surface, which produce an image of the equatorial cross-section of the Fermi surface, as shown in Fig. 2(a).

The remainder of the paper is organized as follows. In Sec.~\ref{sec2} we discuss the topological properties of the line and the point nodes of the $k_z(k_x \pm ik_y)$ state. In Sec.~\ref{sec3} we describe the topology of the point nodes of the $(k_x \pm i k_y)^2$ state. In Sec.~\ref{secnew}, we describe how two candidate chiral pairings can be distinguished through Josephson interferometry. In Sec.~\ref{sec4} and Sec.~\ref{sec5} we respectively consider the surface Andreev bound states for the $m=\pm 1$ and the $m=\pm 2$ pairings, and their experimental consequences. The Sec.~\ref{sec6} and Sec.~\ref{sec7} are respectively devoted to the detailed discussion of anomalous spin and thermal Hall conductivities. Finally we conclude by providing a summary of our results and future directions in Sec.~\ref{sec11}.

\section{Nodal topology of $\mathrm{k}_{\mathrm{z}}(\mathrm{k}_{\mathrm{x}} \pm \mathrm{i} \mathrm{k}_{\mathrm{y}})$ state}\label{sec2}
In this section we consider the topological properties of the nodal excitations of the $k_z(k_x \pm i k_y)$ paired state. It is known that the hidden ordered phase is a compensated semi-metal and possesses multiple electron and hole pockets. Even though the detailed fermiology and the precise nature of the hidden order is still unknown, the details of the participating bands are not too important for the discussion of the topological properties of the paired state. Hence we will ignore the multi-band aspects of the normal and the paired states for simplicity. Our discussions can be easily generalized to the superconductivity of multiple bands.

We begin with the following reduced BCS Hamiltonian operator for the spin singlet, chiral pairing
\begin{eqnarray}\label{hamiltonian1}
\hat{h}_{\mathbf{k}}=\bigg[\xi_{\mathbf{k}}\tau_3+\frac{\Delta_{0}}{k^2_F} \sin \frac{c k_z}{2} \bigg(\sin a \frac{k_x}{2} \tau_1
\pm \sin a\frac{k_y}{2}\tau_2\bigg)\bigg],
\end{eqnarray}
where $\xi_{\mathbf{k}}=\epsilon_{\mathbf{k}}-\mu $ is the kinetic energy of the quasiparticles, measured with respect to the Fermi level $\mu$. The Pauli matrices $\boldsymbol \tau$ operate on the particle-hole indices. The $\pm $ in front of $\tau_2$ respectively correspond to $m= \mp 1$. The pairing amplitudes vanish when (i) $k_z=0, \; \pm 2\pi/c$, for any $k_x$, $k_y$ and (ii) $k_x=k_y=0 \; \pm 2\pi/a $ for any $k_z$. If we consider Fermi pockets around $\Gamma$ or the $Z$ points respectively located at $\mathbf{k}=(0,0,0)$ and $\mathbf{k}=(0,0, \pm 2 \pi/c)$, the pairing amplitude vanishes at the equator of the Fermi surface, giving rise to the line nodes, as shown in Fig. 2(a). Similarly, there are point nodes at the intersections of all the Fermi pockets with the c-axis. For simplicity, we will only consider the Fermi pockets around the $\Gamma$ or the $Z$ points, and the normal state's dispersion will be approximated as $\epsilon_{\mathbf{k}} \approx k^2_z/(2 M_\parallel)+k^2_\perp/(2 M_\perp)$ in the vicinity of these points. For most of our purpose we can also set $M_\perp=M_\parallel=M$, without any loss of generality.

\subsection{Topological invariant of the nodal ring}
The BCS quasiparticles perceive the nodal ring $\mathbf{k}=(k_F \cos \phi_{\mathbf{k}}, k_F \sin \phi_{\mathbf{k}}, 0)$ as a momentum space vortex loop \cite{Beri, Schnyder, Kopnin}, where $\tan \phi_{\mathbf{k}}=k_y/k_x$. For this reason, the quasiparticle wavefunction changes sign while encircling the line node along a closed loop (shown in green in Fig. 2(a)). For describing the topological invariant of the nodal ring, we first notice that the Hamiltonian has the spectral symmetry with respect to the unitary matrix $\mathcal{U}_1=(-\sin \phi_{\mathbf{k}} \tau_1 + \cos \phi_{\mathbf{k}} \tau_2)$, defined as
\begin{equation}\label{symmetry1}
\{\hat{h}_{\mathbf{k}}, \mathcal{U}_1\}=0.\end{equation}
\begin{figure*}[htb]
\centering
\subfigure[]{
\includegraphics[width=8.6cm,height=7cm]{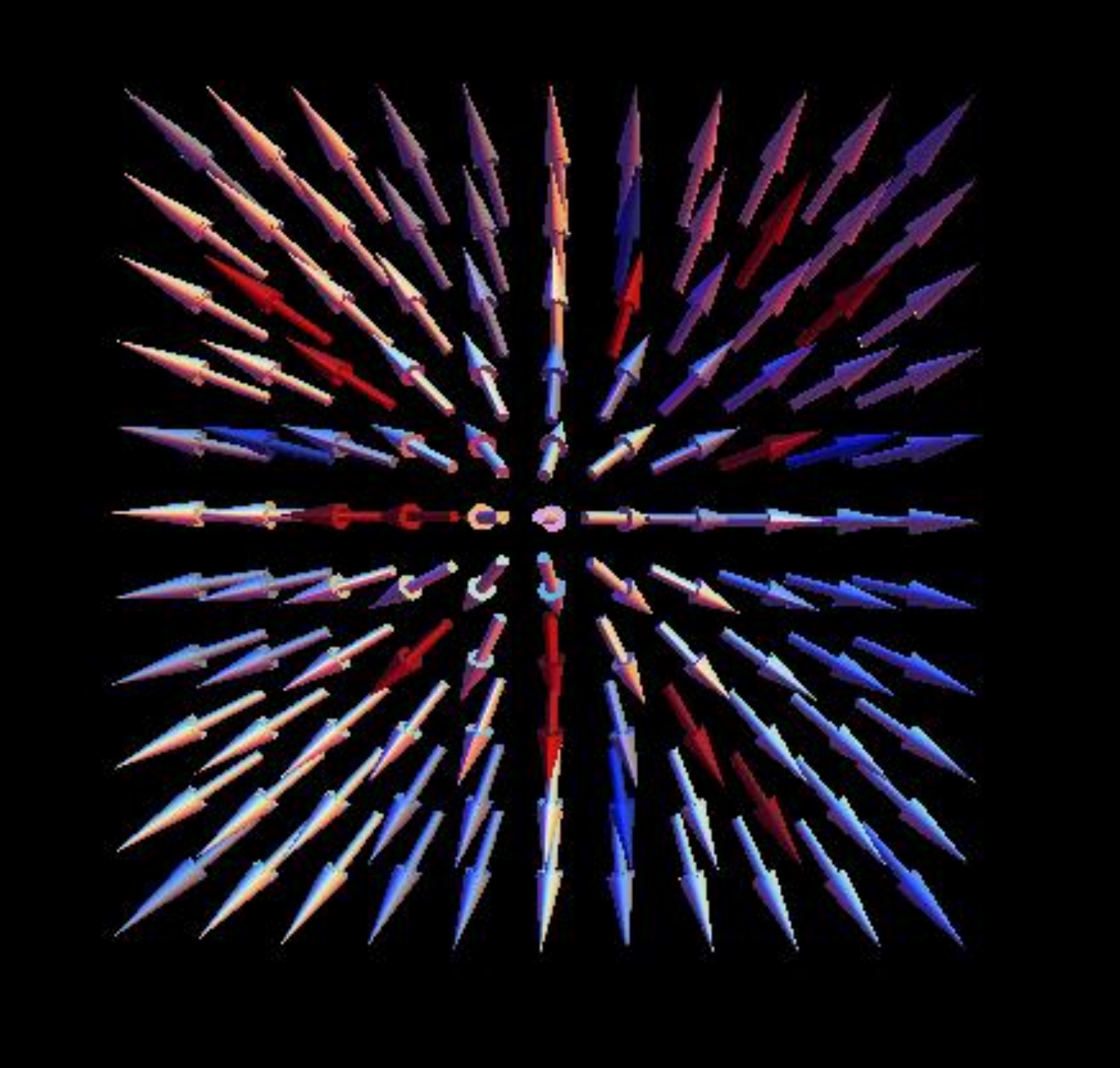}
\label{fig:subfig3a}
}
\subfigure[]{
\includegraphics[width=8.6cm,height=7cm]{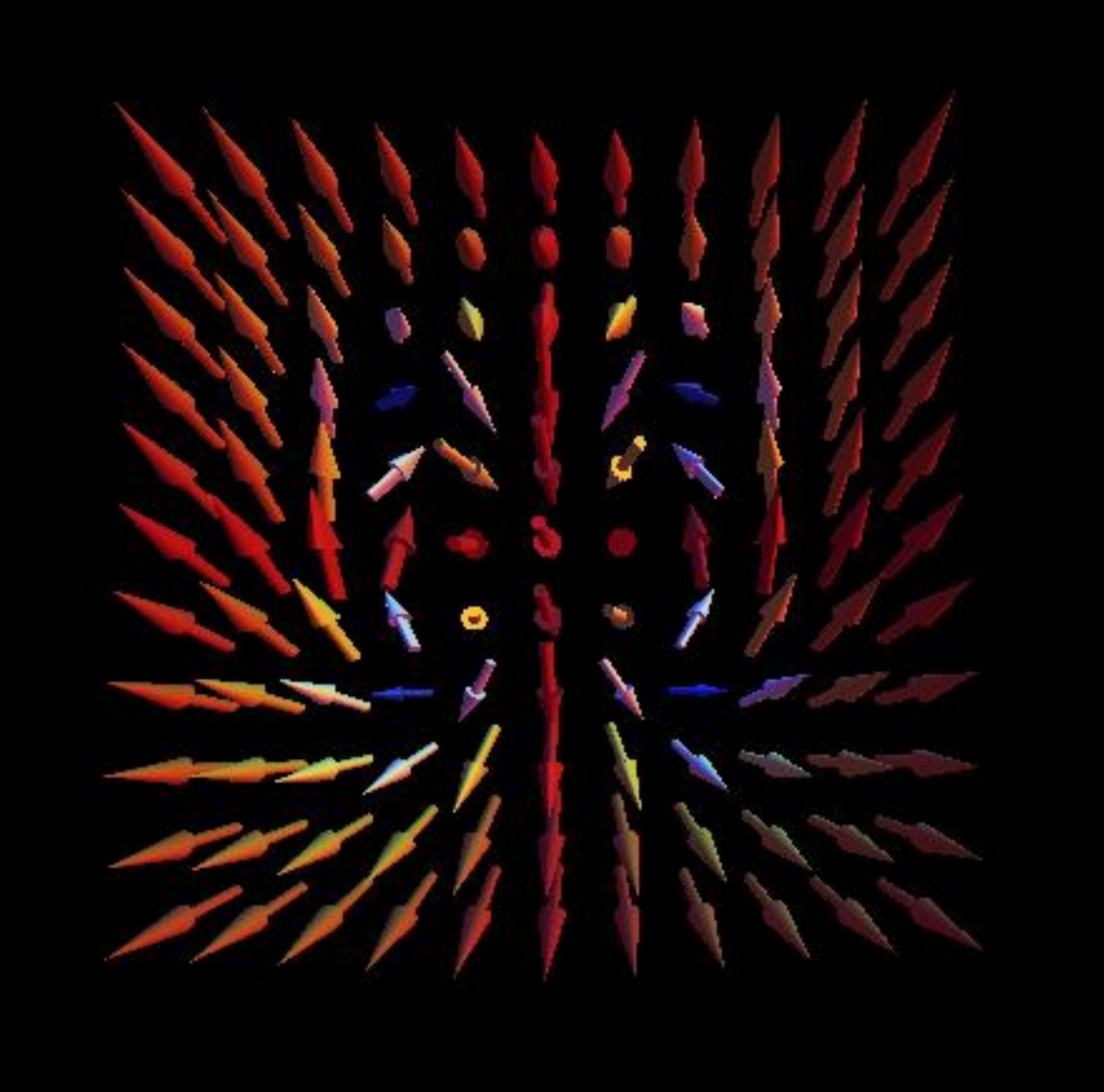}
\label{fig:subfig3b}
}
\label{fig:3}
\caption[]{(a) The unit skyrmion texture of the pseudospin for $k_z(k_x - ik_y)$ pairing, when $k_z \neq 0$ and $-k_F<k_z<k_F$. (b) The double skyrmion texture for $(k_x-ik_y)^2$ pairing, when $-k_F<k_z<k_F$. The skyrmion core size is determined by $\tilde{k}_F=\sqrt{k^2_F-k^2_z}$.}
\end{figure*}Consequently, if $\psi_{\mathbf{k}}$ is an eigenfunction of energy $E$, $\mathcal{U}_1 \psi_{\mathbf{k}}$ is the eigenfunction with energy $-E$. This unitary matrix helps in defining the following topological invariant for any closed loop $\mathcal{C}$, as
\begin{eqnarray}\label{tinvring}
N_{\mathcal{C}}=-\frac{1}{4\pi i} \oint_\mathcal{C} dl \; Tr[(-\sin \phi_{\mathbf{k}} \tau_1 + \cos \phi_{\mathbf{k}} \tau_2) \hat{h}^{-1}_{\mathbf{k}} \partial_{l} \hat{h}_{\mathbf{k}}],
\end{eqnarray}
where $dl$ is the line element along $\mathcal{C}$. For simplicity we can consider $\mathcal{C}$ to be a circle $(k_\perp-k_F)^2+k^2_z=(\delta k)^2$ of radius $\delta k$, and linearize $\hat{h}_{\mathbf{k}}$. After substituting
\begin{eqnarray}
k_x \pm ik_y =(k_F + \delta k \; \sin \alpha_{\mathbf{k}})e^{\pm i\phi_{\mathbf{k}}}, \; k_z=\delta k \cos \alpha_{\mathbf{k}}
\end{eqnarray}
in Eq.~(\ref{hamiltonian1}), we obtain the following linearized Hamiltonian
\begin{eqnarray}
\hat{h}_{\mathbf{k}} \approx \delta k \left[ v_F \sin \alpha_{\mathbf{k}} \; \tau_3 + v_\Delta  \cos \alpha_{\mathbf{k}} \; (\cos \phi_{\mathbf{k}} \; \tau_1 + \sin \phi_{\mathbf{k}} \; \tau_2)\right] \nonumber \\
\end{eqnarray}
where $v_F=k_F/M$, $v_\Delta=\Delta_0/k_F$. Substituting this linearized form in Eq.~(\ref{tinvring}), we find
\begin{eqnarray}
N_{\mathcal{C}}=\frac{v_F v_\Delta}{2\pi} \int^{2\pi}_0 \frac{d\alpha_{\mathbf{k}}}{v^2_F \sin^2 \alpha_{\mathbf{k}} + v^2_\Delta \cos^2 \alpha_{\mathbf{k}}}=1.
\end{eqnarray}
Due to the existence of this topological invariant in the bulk, the line node will give rise to zero energy Andreev bound states on the $(0,0,1)$ surface. The emergence of such surface states are discussed in detail in Sec.~\ref{sec4}.

\subsection{Topological invariant of the nodal points}
\begin{figure*}[htb]
\includegraphics[width=14cm,height=14cm]{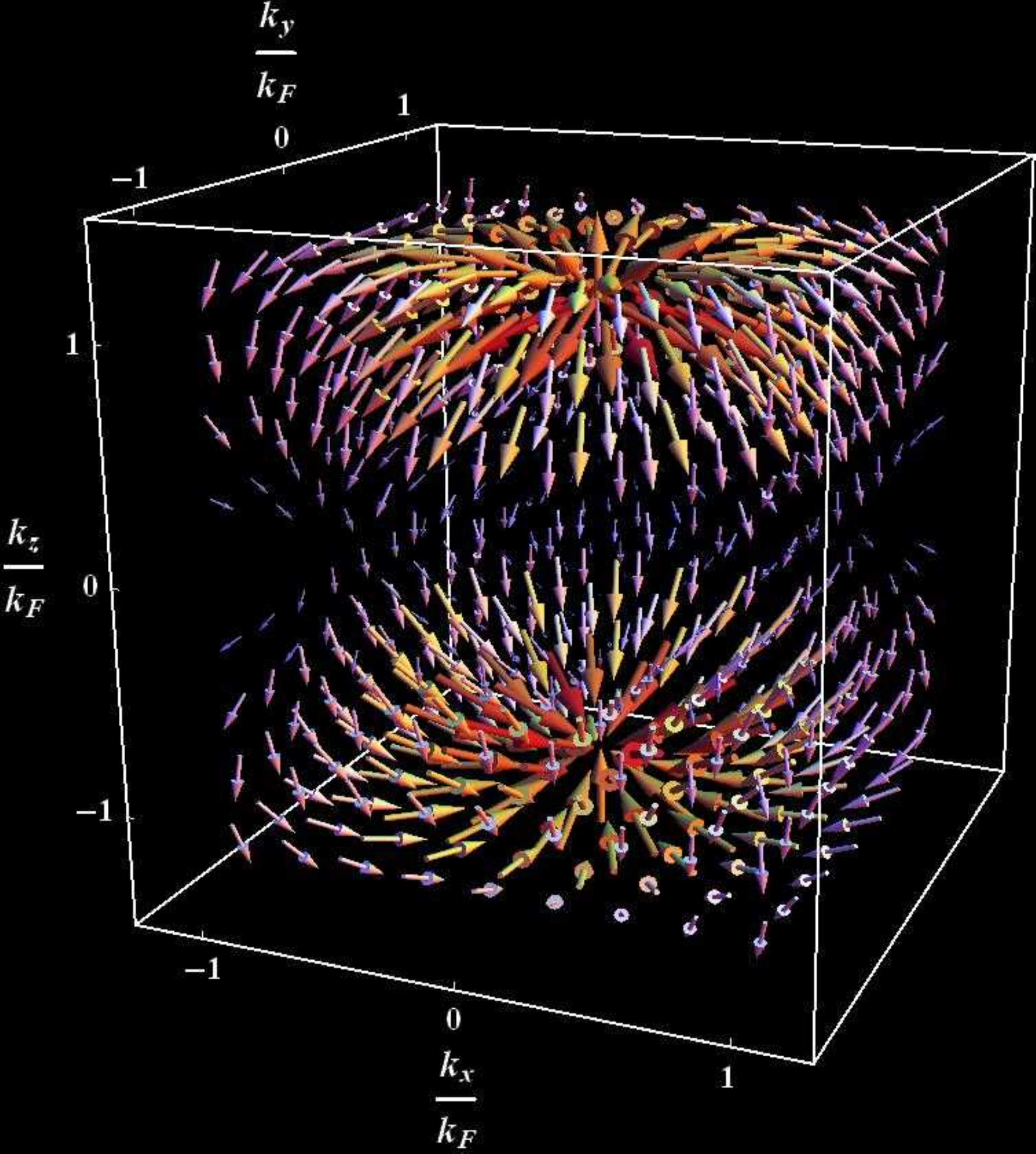}
\caption{(Color online)A vector-field plot of the quasiparticle's Berry curvature for $\Delta_{\mathbf{k}}= \Delta_{0}k_z(k_x - ik_y)/k^2_F$. The point nodes at the Fermi surface poles appear as the unit charge (anti)monopole of the Berry curvature and give rise to the linearly dispersing Weyl fermions. The Berry curvature is diminished as the equator of the Fermi surface is approached, due to the presence of the line node.  }
\label{fig:4}
\end{figure*}
The discussion of the topological aspects of the nodal points can be facilitated by rewriting the Hamiltonian as $\hat{h}_{\mathbf{k}}=\mathbf{N}_\mathbf{k} \cdot \boldsymbol \tau$, where $\mathbf{N}_{\mathbf{k}}=(\Delta_{1,\mathbf{k}},\Delta_{2,\mathbf{k}},\xi_{\mathbf{k}})$ is a pseudospin vector, and $\Delta_{1,\mathbf{k}}$, $\Delta_{2,\mathbf{k}}$ are the real and the imaginary parts of the pairing amplitude. For any $-k_F<k_z<k_F$, where $k_F=\sqrt{2M\mu}$ is the Fermi momentum, we have an effective two dimensional problem with a modified Fermi energy $\mu-k^2_z/(2M)$. If the spectrum is fully gapped ($|\mathbf{N}_{\mathbf{k}}| \neq 0$), the unit pseudospin vector $\mathbf{n}_{\mathbf{k}}=\mathbf{N}_{\mathbf{k}}/|\mathbf{N}_{\mathbf{k}}|$ can possess a skyrmion texture. A skyrmion texture is described by
\begin{equation}
\mathbf{n}=\bigg [ \sin \left(f(k_\perp) \right) \cos m \phi, \sin \left(f(k_\perp)\right) \sin m \phi, \cos \left(f(k_\perp)\right)\bigg],
\end{equation}
where $f(k_\perp=0)=\pi$, $f(k_\perp=\infty)=0$, $k_\perp=\sqrt{k^2_x+k^2_y}$, and $\phi_k=\tan^{-1}(k_x/k_y)$. These textures are characterized by the integer topological invariant
\begin{eqnarray}
W_{s}[\mathbf{n}_{\mathbf{k}}]=\int \frac{d^2 k_\perp}{4\pi} \; \mathbf{n}_{\mathbf{k}} \cdot \bigg(\frac{\partial \mathbf{n}_{\mathbf{k}}}{\partial k_x} \times \frac{\partial \mathbf{n}_{\mathbf{k}}}{\partial k_y}\bigg)
\end{eqnarray}
For $k_z(k_x \pm ik_y)$ pairing, the underlying Fermi surface is fully gapped when $-k_F<k_z<k_F$ and $k_z \neq 0$, and
\begin{equation}\label{sk1}
W_s[\mathbf{n}]=m=\pm 1.
 \end{equation}
Such a unit skyrmion texture is shown in Fig.~\ref{fig:subfig3a}. The Fermi surface poles at $\mathbf{k}=(0,0,\pm k_F)$ appear as the special singular points, where the skyrmion number changes from $m=\pm 1$ to zero. Such singular point defects of the unit vector field are known as the (anti)hedgehogs, and are characterized by the following topological invariant
\begin{figure*}[htb]
\includegraphics[width=14cm,height=14cm]{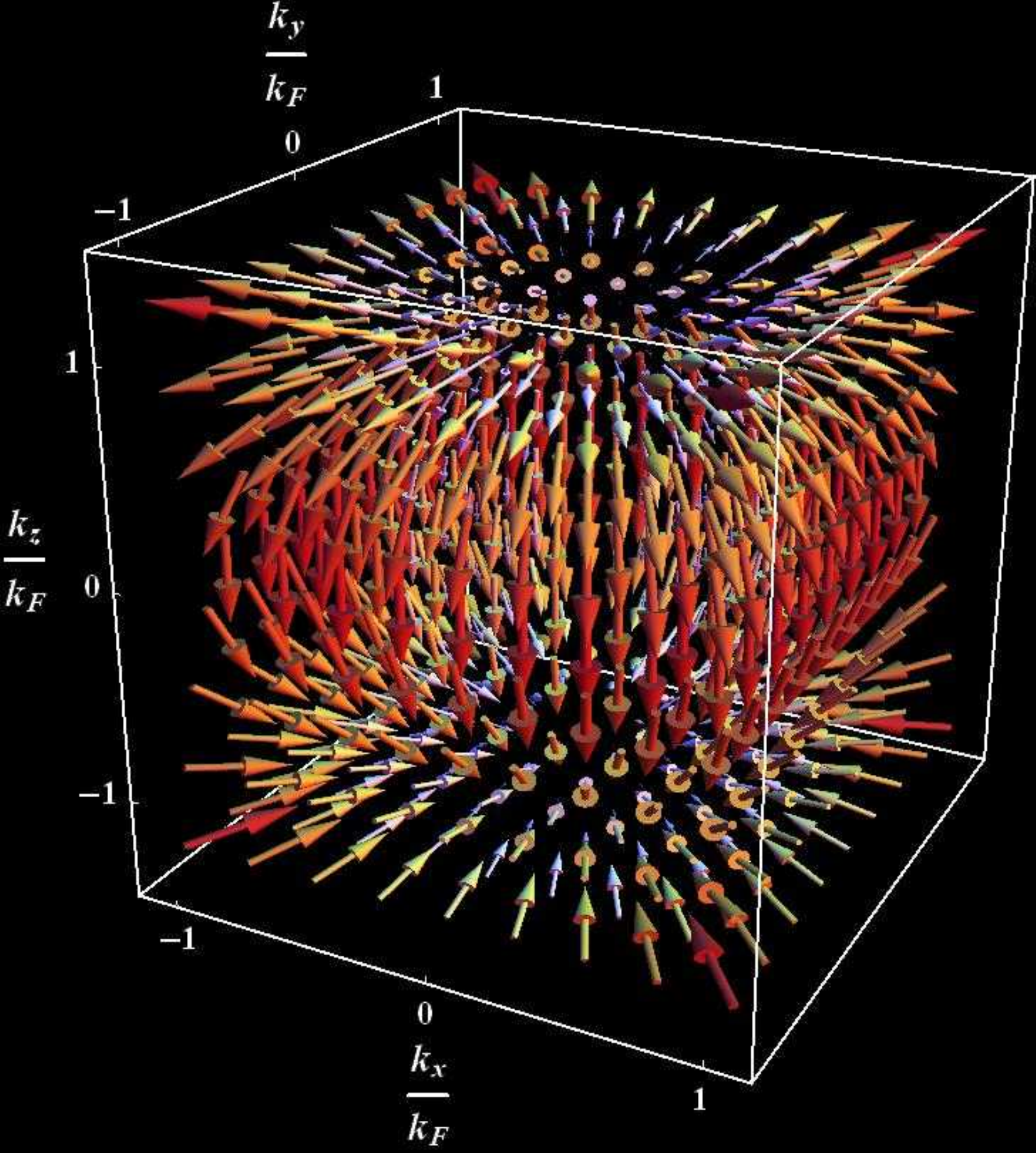}
\caption{(Color online)A vector-field plot of the quasiparticle's Berry curvature for $\Delta_{\mathbf{k}}= \Delta_{0}(k_x - ik_y)^2/k^2_F$. The point nodes at the Fermi surface poles act as the double (anti)monopole of the Berry curvature and generalize the notion of Weyl fermions. }
\label{fig:5}
\end{figure*}
\begin{equation}
W_{h}[\mathbf{n}_{\mathbf{k}}]=\frac{1}{4\pi} \; \int^\pi_0 \; d \theta_{\mathbf{k}} \; \int^{2\pi}_0 \; d\phi_{\mathbf{k}} \; \mathbf{n}_{\mathbf{k}} \cdot \bigg(\frac{\partial \mathbf{n}_{\mathbf{k}}}{\partial \theta_{\mathbf{k}}} \times \frac{\partial \mathbf{n}_{\mathbf{k}}}{\partial \phi_{\mathbf{k}}}\bigg),
\end{equation}
where we are considering a sphere of radius $\delta k$ around the singular points, and $\theta_\mathbf{k}$ and $\phi_{\mathbf{k}}$ are the polar and the azimuthal angles. The point nodes of $k_z(k_x \pm i k_y)$ pairing correspond to $W_h=\pm 1$ and describe the coveted Weyl fermions. When, $m=+1$, the right and the left handed fermions are respectively located at $(0,0,\mp k_F)$. The locations of the Weyl points are interchanged for $m=-1$. By linearizing the Hamiltonian in Eq.~(\ref{hamiltonian1}) for $m=-1$ with respect to the momenta $(0,0,\pm k_F)$ we obtain the Weyl Hamiltonians
\begin{equation}\label{Weyl}
H_{R/L}=\pm \bigg[v_F (k_z \mp k_F)\tau_3+v_{\Delta}\left( k_x \tau_1 + k_y \tau_2\right) \bigg],
\end{equation}
where $v_F=k_F/M$, $v_{\Delta}=\Delta/k_F$. If we define the following parametrization $\delta \mathbf{k}_\pm=(\mathbf{k}\pm k_F\hat{z})=\delta k(\sin \theta \cos \phi, \sin \theta \sin \phi, \cos \theta)$ and compute the hedgehog invariant, we indeed find $W_h=\pm 1$ respectively for the right and the left handed Weyl fermions.

The topological invariant for the nodal points can be given more precise meaning in terms of the Berry curvature, which is also applicable for a model defined on a lattice, where the notion of the skyrmion number should be replaced by the Chern number. From the Bloch wave-functions $\psi_{\mathbf{k},n}$ of the BCS quasiparticles, where $n=\pm$ are the band indices, we can obtain the fictitious vector potential $\mathcal{A}_{\mathbf{k},n}=i \langle \psi_{\mathbf{k},n} | \nabla_{\mathbf{k}} | \psi_{\mathbf{k},n} \rangle$ and the Berry curvature $\mathbf{\Omega}_{\mathbf{k},n}=\nabla_{\mathbf{k}} \times \mathcal{A}_{\mathbf{k},n}$. The Berry curvature can be expressed in terms of the unit vector $\mathbf{n}_{\mathbf{k}}$ according to
\begin{eqnarray}\label{Berry}
\Omega_{\mathbf{k},n,a}=\frac{(-1)^n}{4} \; \epsilon_{abc} \; \cdot \bigg[ \frac{\partial \mathbf{n}_{\mathbf{k}}}{\partial k_b}\times \frac{\partial \mathbf{n}_{\mathbf{k}}}{\partial k_c}\bigg].
\end{eqnarray}For $k_z(k_x \pm ik_y)$ pairing, three components of the Berry curvature are given by
\begin{eqnarray}
\Omega_{\mathbf{k},n,x}= \frac{(-1)^{n+1}\mathrm{sgn}(m) \; \Delta^2_0\;k_zk_x(k^2_z-k^2_\perp+k^2_F)}{2\mu^2 \; \left[(k^2-k^2_F)^2+\frac{\Delta^2_0}{\mu^2}k^2_zk^2_\perp\right]^{\frac{3}{2}}},
\end{eqnarray}
\begin{eqnarray}
\Omega_{\mathbf{k},n,y}= \frac{(-1)^{n+1}\mathrm{sgn}(m) \; \Delta^2_0 \; k_zk_y(k^2_z-k^2_\perp+k^2_F)}{2\mu^2 \; \left[(k^2-k^2_F)^2+\frac{\Delta^2_0}{\mu^2}k^2_zk^2_\perp\right]^{\frac{3}{2}}}, \\
\Omega_{\mathbf{k},n,z}= \frac{(-1)^{n+1}\mathrm{sgn}(m) \; \Delta^2_0 \; k^2_z(k^2_z-k^2_\perp-k^2_F)}{2\mu^2 \; \left[(k^2-k^2_F)^2+\frac{\Delta^2_0}{\mu^2}k^2_zk^2_\perp\right]^{\frac{3}{2}}}.
\end{eqnarray}From these expressions we can immediately see that the right and the left Weyl points respectively act as the unit monopole and the unit antimonopole of the Berry curvature. For illustrative purpose, we have plotted the Berry curvature for the $+$ band and $m=-1$ in Fig.~\ref{fig:4}. Notice that $\Omega_{\mathbf{k},n,x/y}$ are odd functions of $k_z$ and $k_{x/y}$, and consequently the flux through the $ca$ and the $cb$ planes vanish. However, the flux through the $ab$ plane, which is perpendicular to the nodal separation equals $2\pi$ when the plane is located between the nodal points. The Berry flux is $2\pi$ times the Chern number or the Skyrmion number in Eq.~(\ref{sk1}). In Sec.~\ref{sec4}, we will show this flux to be responsible for giving rise to the chiral surface states on the planes, whose normals are perpendicular to $c$-axis, and also for the existence of anomalous spin and thermal Hall conductivities.

\section{Nodal topology of $(\mathrm{k}_{\mathrm{x}} \pm \mathrm{i} \mathrm{k}_{\mathrm{y}})^2$ state}\label{sec3}
The $\Delta^{\prime}(\mathbf{k})$ in Eq.~(\ref{pairing2}) vanishes when (i) $k_z= \pm \pi/c$, for any $k_x$, $k_y$ and (ii) $k_x=k_y=0$ for any $k_z$. Consequently, any Fermi pocket around $k_z= \pm \pi/c$ will possess line nodes, and the intersections of the $c$-axis and the Fermi pockets lead to the point nodes (as shown in Fig.~2(b)). For simplicity, we will again restrict ourselves to the Fermi pockets around $\Gamma$ or $Z$ points, and the pertinent BCS Hamiltonian
becomes
\begin{eqnarray}\label{hamiltonian2}
\hat{h}_{\mathbf{k}}=\bigg[\xi_{\mathbf{k}}\tau_3+\frac{\Delta_{0}}{k^2_F} \left \{(k^2_x-k^2_y) \tau_1
\pm 2k_xk_y\tau_2\right \} \bigg],
\end{eqnarray} For $-k_F< k_z <k_F$, the unit vector $\mathbf{n}_{\mathbf{k}}$ now describes a pseudospin skyrmion texture, with $W_{sk}=\pm 2$, as shown in Fig.~\ref{fig:subfig3b}.

The Fermi points correspond to charge two (anti)hedgehogs. If we expand the BCS Hamiltonian for $m=- 2$ pairing around the Fermi points, we obtain
\begin{eqnarray}\label{doubleWeyl}
H_{R/L}=\bigg[\left \{ \pm v_F(k_z \mp k_F)+\frac{(k_z \mp k_F)^2}{2M}+\frac{k^2_\perp}{2M} \right \} \tau_3 \nonumber \\ +\frac{\Delta}{k^2_F}\left((k^2_x-k^2_y) \tau_1+2k_x k_y \tau_2 \right)\bigg]
\end{eqnarray}
Therefore, the low energy nodal quasiparticles possess linear and quadratic dispersions along the $c$ axis and in the $ab$ plane respectively. This anisotropic dispersion is responsible for giving rise to the linear density of states at low energies, which can be obtained from
\begin{eqnarray}
D(\epsilon)=2 \int \frac{d^3 k}{(2\pi)^3} \delta \bigg(\epsilon-\bigg[\frac{(k^2-k^2_F)^2}{4M^2} \nonumber \\ +\frac{\Delta^2_0}{k^4_F} k^4 \sin^4 \theta_{\mathbf{k}} \bigg]^{\frac{1}{2}} \bigg).
\end{eqnarray}
The integral over $k$ is performed in the range $k^2_F-2M\Omega_D<k^2<k^2_F+2M\Omega_D$, where $\Omega_D$ is the Debye cut-off for the pairing interaction. After substituting $(k^2-k^2_F)/(2M)=x$ and noting that $x \ll k^2_F$, we find $\sin^4 \theta_{\mathbf{k}}=(\epsilon^2-x^2)/\Delta^2_0$ and two constraints (i) $- \epsilon <x<  \epsilon$, (ii) $x^2> \epsilon^2-\Delta^2_0$. For low energies, $\epsilon \ll \Delta_0$, both constraints are automatically satisfied. After some simple algebra we obtain
\begin{eqnarray}\label{DOSdoubleWeyl}
D(\epsilon)=  \frac{Mk_F|\epsilon|}{\pi^2\Delta_0} \bigg[\Theta(\Delta^2-\epsilon^2) \int^{\pi/2}_0 \frac{ dy}{\sqrt{1-\frac{\epsilon}{\Delta_0}\cos y}}\nonumber \\ + \Theta(\Omega^2_D-\epsilon^2)\Theta(\epsilon^2-\Delta^2)\int^{\pi/2}_{y_0} \frac{ dy}{\sqrt{1-\frac{\epsilon}{\Delta_0}\cos y}}\bigg],
\end{eqnarray}
where $\sin y_0=\sqrt{1-\Delta^2_0/\epsilon^2}$, and $\Theta(x)$ is the Heaviside step function. For $\epsilon \ll \Delta_0$ the first integral $\sim \pi/2$ and we obtain linear density of states. For this reason, even in the absence of the line nodes, $(k_x \pm ik_y)^2$ state can adequately describe the low temperature thermodynamic properties of superconducting URu$_2$Si$_2$.

The Berry curvature for this pairing are given by
\begin{eqnarray}
&&\Omega_{\mathbf{k},n,x}= \frac{(-1)^{n+1}\mathrm{sgn}(m) \; 2\Delta^2_0 \; k_zk_xk^2_\perp}{\mu^2 \; \left[(k^2-k^2_F)^2+\frac{\Delta^2_0}{\mu^2}k^4_\perp\right]^{\frac{3}{2}}},\\
&&\Omega_{\mathbf{k},n,y}=\frac{(-1)^{n+1}\mathrm{sgn}(m) \; 2\Delta^2_0 \; k_zk_yk^2_\perp}{\mu^2 \: \left[(k^2-k^2_F)^2+\frac{\Delta^2_0}{\mu^2}k^4_\perp\right]^{\frac{3}{2}}},\\
&&\Omega_{\mathbf{k},n,z}=\frac{(-1)^{n+1}\mathrm{sgn}(m) \; 2\Delta^2_0 \; (k^2_z-k^2_F)k^2_\perp}{\mu^2 \left[(k^2-k^2_F)^2+\frac{\Delta^2_0}{\mu^2}k^4_\perp\right]^{\frac{3}{2}}}.
\end{eqnarray}
The Berry curvatures for the $m=-2$ and $n=+$ band are shown in Fig.~\ref{fig:4}. Using the above expressions we can explicitly demonstrate that the Fermi points act as the double Berry (anti)monopoles. As before, there is no net flux through the $ca$ and the $cb$ planes. But, the flux through the $ab$ plane equals $4\pi$. Since the topological invariant for this pairing is two times bigger than the one for the $k_z(k_x \pm i k_y)$ pairing, there will be two times as many chiral surface states, as shown in Sec.~\ref{sec4}.

\section{Bulk Topological invariant and Josephson interferometry}\label{secnew}
Given that we have two possible topologically distinct chiral pairings, a natural question is if we can experimentally distinguish them. The bulk invariant $m$ can be determined through the Josephson interferometry, which can provide the relevant phase sensitive information regarding the pairing symmetry\cite{Ginsberg,LiuMaeno,StrandMaeno,Halperin}. For $k_z(k_x \pm ik_y)$ and $(k_x \pm i k_y)^2$ pairings, a $\pi/2$ rotation about $c$-axis respectively change the pairing amplitudes by $e^{\i \pi/2}$ and $e^{i \pi}$. This phase shift can be measured in a corner junction set up, where both $a$ and $b$ axes are used to define the corner. This method has already been used for finding the pairing symmetry of the B-phase in superconducting UPt$_3$, where the observed $\pi$ phase shift supports the triplet $k_z(k_x \pm i k_y)^2$ state (chiral f-wave, $Y^{\pm 2}_3(\hat{\Omega}$)), and rules out the possible singlet $k_z(k_x \pm ik_y)$ state \cite{Halperin}. Therefore, future corner Josephson junction and SQUID measurements will also be instrumental in settling the pairing symmetry of URu$_2$Si$_2$. In addition the surface measurements such as ARPES and Fourier transformed STM can also elucidate the topological nature of the pairing by directly probing the surface Andreev bound states. For this reason, we present a detailed calculation of the surface Andreev bound states for both pairings.

\section{Surface Andreev Bound States for $k_z(k_x \pm ik_y)$ state}\label{sec4}
Earlier we have noted the presence of two distinct nodal topologies of the $k_z(k_x \pm i k_y)$ state, associated with the nodal ring and the nodal points. When the superconductor is terminated at a surface with the vacuum, the topological invariants jump to zero, which give rise to gapless surface Andreev bound states.

\subsection{Dispersionless bound states on $(0,0,1)$ surface}

Let us first consider a boundary at $z=0$, such that $z<0$ and $z>0$ regions are respectively occupied by the superconductor and the vacuum. The two component spinor wave function $\psi^T=(u,v)$ for zero energy surface Andreev bound state satisfies the following differential equations
\begin{eqnarray}\label{surface1}
&&\left[\left(-\partial^2_z+k^2_{\perp}-k^2_F\right)\tau_3 -\frac{i\Delta}{\mu}\; k_\perp \left(\cos \phi \tau_1 + \sin \phi \tau_2\right)\partial_z\right]\psi \nonumber \\&&=0,
\end{eqnarray}
and the boundary conditions $\psi(z=0)=\psi(z\to -\infty)=0$. Due to the spectral symmetry of the Hamiltonian with respect to the matrix $\mathcal{U}_1$ in Eq.~(\ref{symmetry1}), the zero energy states have to be eigenstates of $\mathcal{U}_1$ with eigenvalue $+1$ or $-1$. This is tied to an index theorem, which mandates
\begin{equation}\label{index1}
\langle \psi| \mathcal{U}_1 |\psi \rangle=n_+-n_-=N_{\mathcal{C}}(z=+\infty)-N_{\mathcal{C}}(z=-\infty),
\end{equation}
where $n_\pm$ respectively represent the number of zero energy states with eigenvalues $\pm 1$. Therefore, we substitute
\begin{equation}\label{zeromode1}
\psi_\pm(x,y,z)=f_\pm(z) \; e^{i(k_xx+k_yy)} \; \left(\begin{array}{c}
e^{-\frac{i}{2}\left(\phi_{\mathbf{k}}\pm \frac{\pi}{2}\right)} \\
e^{\frac{i}{2}\left(\phi_{\mathbf{k}}\pm \frac{\pi}{2}\right)} \end{array} \right )
\end{equation}
in Eq.~(\ref{surface1}), and $f_\pm(z) \sim e^{\lambda_\pm z}$, leading to the following secular equations for $\lambda_\pm$
\begin{eqnarray}
\lambda^2_\pm \mp \frac{\Delta_0}{\mu} k_\perp \lambda_\pm -(k^2_\perp-k^2_F)=0.
\end{eqnarray}
The secular equations have the following solutions
\begin{eqnarray}
\lambda_{+,j}=\frac{\Delta_0}{2\mu}k_\perp +(-1)^j \sqrt{\left(1+\frac{\Delta^2_0}{4\mu^2}k^2_\perp \right)-k^2_F}, \\
\lambda_{-,j}=-\frac{\Delta_0}{2\mu}k_\perp +(-1)^j \sqrt{\left(1+\frac{\Delta^2_0}{4\mu^2}k^2_\perp \right)-k^2_F},
\end{eqnarray}
where $j=1,2$. The boundary condition $\psi(z\to -\infty)=0$ can only be compatible with $\Psi_+(z) \sim e^{\lambda_{+,j}z}$, which immediately sets $n_-=0$ in Eq.~(\ref{index1}). In order to satisfy the boundary condition at $z=0$, we require $f_+(z)=A_1 \; \sum_j (-1)^j \; e^{\lambda_{+,j}z}$, and obtain
\begin{eqnarray}
&&f_+(z)=\Theta(-z) \; \left[\frac{\Delta_0k_\perp}{\mu}\frac{k^2_F-k^2_\perp}{k^2_F-\left(1+\frac{\Delta^2_0}{4\mu^2}k^2_\perp \right)}\right]^{\frac{1}{2}}\nonumber \\ && \times e^{\frac{\Delta_0k_\perp z}{2\mu}}\; \sinh \left[z\sqrt{\left(1+\frac{\Delta^2_0}{4\mu^2}k^2_\perp \right)-k^2_F}\right],
\end{eqnarray}
which is only normalizable for $k_F>k_\perp$ and $k_\perp \neq 0$. Therefore, the dispersionless zero energy bound states on the $(0,0,1)$ surface produce an image of the equatorial cross-section of the Fermi surface, as shown in Fig.~2(a). Due to the spin-singlet nature of the pairing, the bound states are two fold spin-degenerate. These zero energy states possess divergent density of states, and can give rise to zero bias peak in tunneling measurements \cite{Beri, Schnyder, Kopnin}. The Zeeman coupling due to an external magnetic field lifts the spin degeneracy and also makes these bound states gapped.
\begin{figure*}[htb]
\centering
\subfigure[]{
\includegraphics[width=8.6cm,height=7cm]{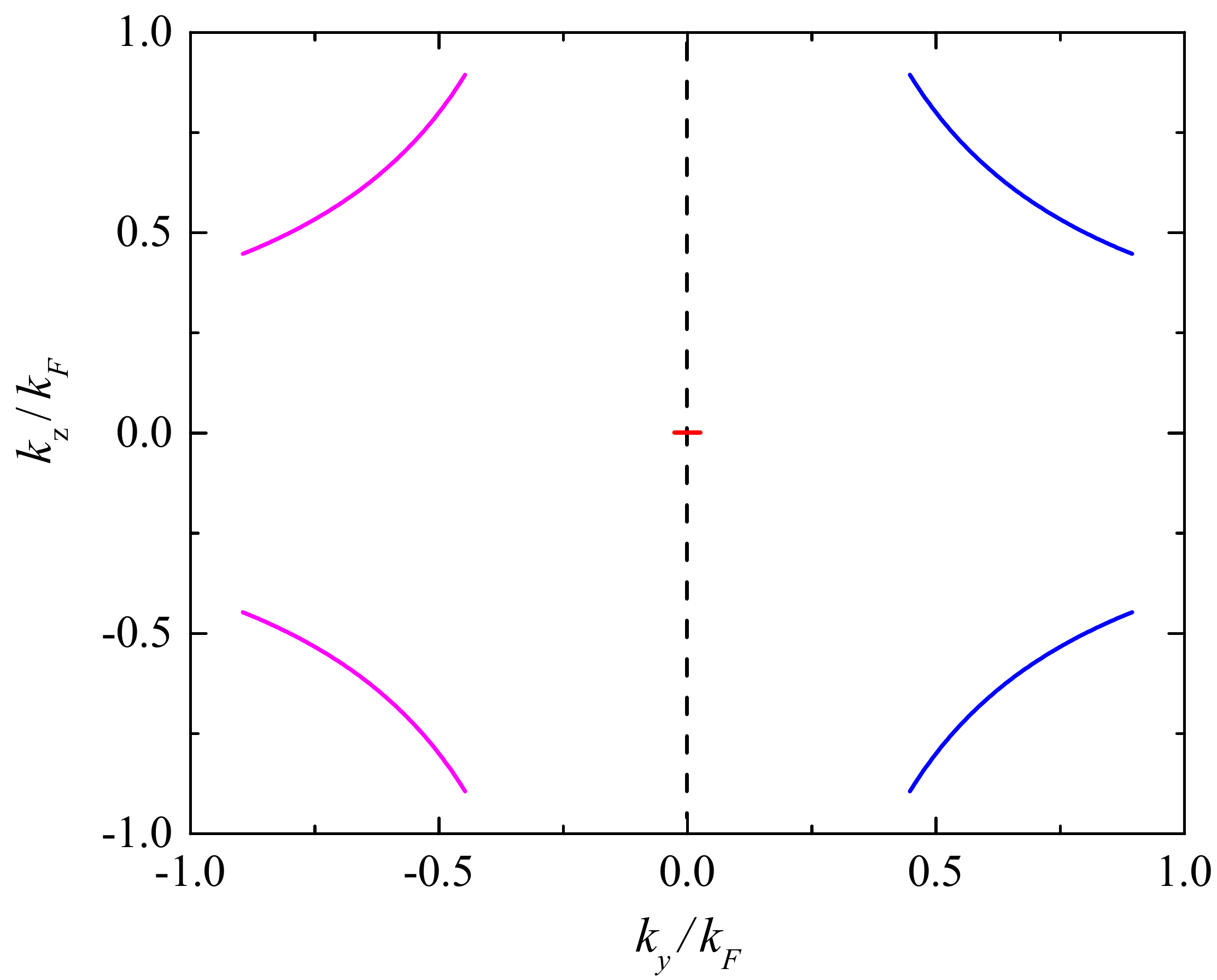}
\label{fig:subfig8a}
}
\subfigure[]{
\includegraphics[width=8.6cm,height=7cm]{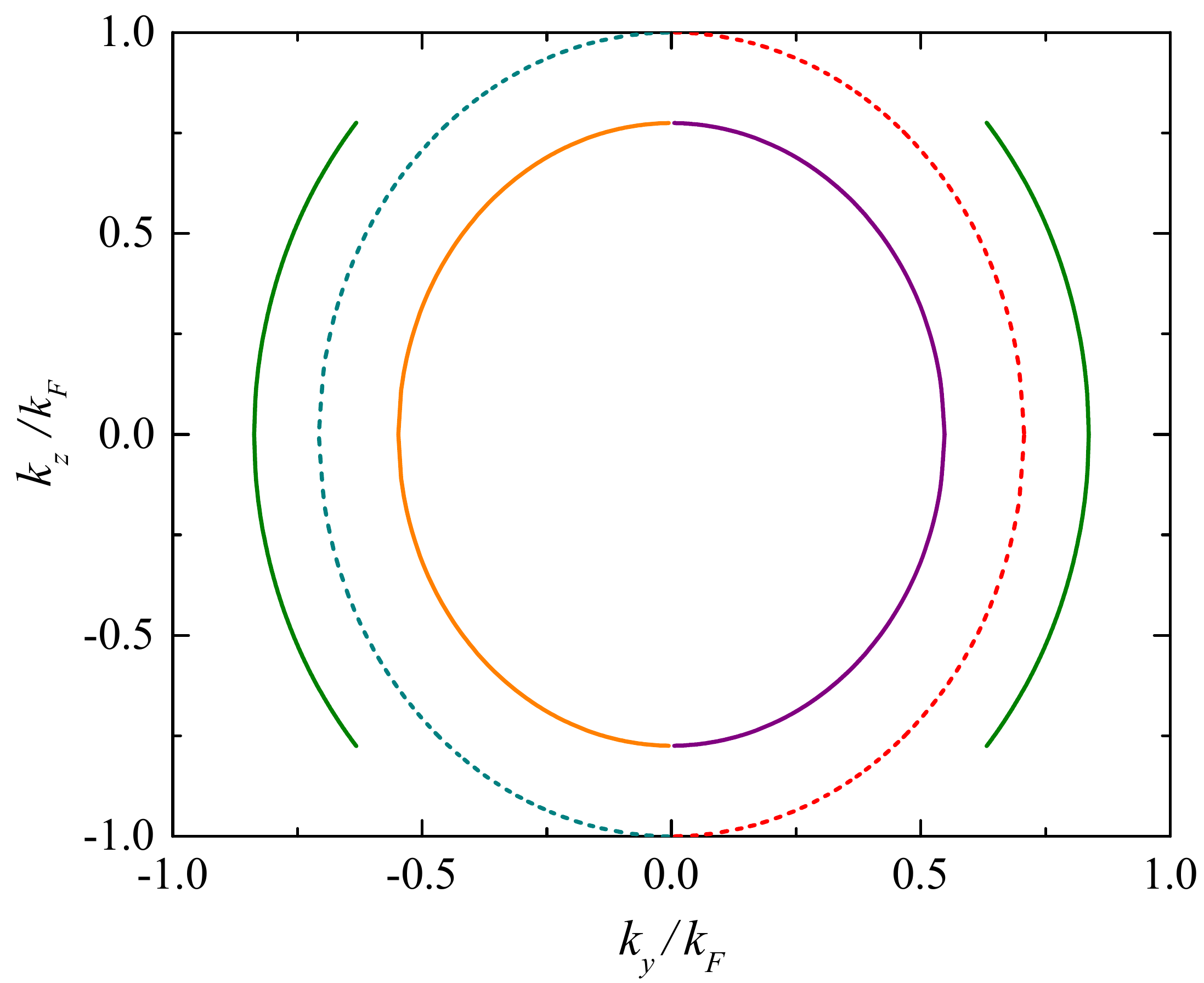}
\label{fig:subfig8b}
}
\label{fig:8}
\caption[]{(a) The dashed line represents the spin degenerate Majorana arcs of the $k_z(k_x \pm ik_y)$ pairing, on $(1,0,0)$ surface, and the absence of the bound state at $k_z=0$ is indicated by the red slash. A uniform Zeeman coupling leads to spin non-degenerate Majorana arcs, which are shown as solid lines. (b) The dashed lines represent two spin degenerate Majorana arcs for $(k_x \pm i k_y)^2$ pairing on $(1,0,0)$ surface. A uniform Zeeman coupling leads to four spin non-degenerate Majorana arcs, as shown by the solid lines. There is an overall decrease in the number of chiral modes due to the Zeeman coupling, which causes a reduction of anomalous spin and thermal Hall conductivities.}
\end{figure*}
\subsection{Chiral bound states on $(1,0,0)$ surface}
Now consider a boundary in the x-direction, such that $x<0$ and $x>0$ regions are respectively occupied by the superconductor and the vacuum. The spinor wave-function now satisfies the differential equations
\begin{eqnarray}
&&\left[\left(-\partial^2_x+k^2_{y}+k^2_z-k^2_F\right)\tau_3 +\frac{\Delta_0 k_z}{\mu}\;\left(-i\partial_x \tau_1 +k_y \tau_2\right)\right]\psi \nonumber \\ && \: \: \: \: \: \: \: \: \: \: \: =2mE \psi,
\end{eqnarray}
and the boundary conditions $\psi(0)=\psi(x\to -\infty)=0$. Notice that we are working with $m=-1$ pairing. When $k_y=0$, the Hamiltonian operator anticommutes with $\tau_2$. Therefore, the zero energy states have to be eigenstates of $\tau_2$ and satisfy a similar index theorem as in Eq.~(\ref{index1}). The decay of the bound state wavefunction is controlled by $\Delta_0 k_z/\mu$. Hence, the bound states are eigenstates of $\tau_2$ with eigenvalues $\mathrm{sgn} (k_z)$ for $k_z \neq 0$. As the gap vanishes at $k_z=0$, there can be no bound state on the $(1,0,0)$ surface corresponding to $k_z=0$. This is illustrated in the Fig.~2(a), and Fig.~\ref{fig:subfig8a} as a discontinuity of the Fermi arc. The precise statement of the index theorem is given by
\begin{eqnarray}
\langle \Psi| \tau_2 |\Psi \rangle (k_z,k_y)=n_+-n_-= \mathrm{sgn}(k_z),
\end{eqnarray}
where $n_\pm$ are respectively the number of zero energy bound states which are eigenfunctions of $\tau_2$ with eigenvalues $\pm 1$.

The eigenfunction and the dispersion relation of the bound states are respectively described by
\begin{eqnarray}
&&\psi(x,y,z)=u(x) \; e^{(ik_y y+k_z z)}\left(\begin{array}{c}
1 \\
 i \; \mathrm{sgn}(k_z)  \end{array} \right ), \\
 && E=+ \frac{\Delta_0}{k^2_F}|k_z| k_y,
\end{eqnarray}
which shows the existence of chiral dispersion along the $y$ direction, and $u(x)$ satisfies the differential equation
\begin{eqnarray}
\partial^2_x u - \frac{\Delta}{\mu} |k_z| \partial_x u -(k^2_y+k^2_z-k^2_F)u=0.
\end{eqnarray} After substituting $u(x)=A_2 \; \sum_j (-1)^j \; e^{\lambda_{+,j}x}$ where $j=1,2$, and $A_2$ is a normalization constant, we find
\begin{eqnarray}
\lambda= \frac{\Delta}{2\mu} |k_z| \pm \sqrt{k^2_z\left(1+\frac{\Delta^2}{4\mu^2}\right)+k^2_y-k^2_F}.
\end{eqnarray}
Therefore the spinor wave-function is given by
\begin{eqnarray}\label{wafn}
&&\psi(x,y,z)= A_2 \; \left(\begin{array}{c}
1 \\
 i \; \mathrm{sgn}(k_z)  \end{array} \right ) \; \Theta(-x) \; e^{i(k_y y+k_z z)} \nonumber \\ && \times  e^{\left(\frac{\Delta |k_z|}{2\mu} \; x\right)} \; \sinh \bigg[x \sqrt{k^2_z\left(1+\frac{\Delta^2}{4\mu^2}\right)+k^2_y-k^2_F}\bigg] , \nonumber \\
\end{eqnarray}
and
\begin{eqnarray}
A_2=\left[\frac{\Delta_0|k_z|}{\mu}\frac{k^2_F-k^2_y-k^2_z}{k^2_F-k^2_y-\left(1+\frac{\Delta^2_0}{4\mu^2} \right)k^2_z}\right]^{\frac{1}{2}}.
\end{eqnarray} Only when $k^2_z+k^2_y < k^2_F$ and $k_z \neq 0$, the wavefunction is normalizable. These bound state solutions are chiral Majorana fermions and their energies vanish along the line $k_y=0$ for any $-k_F<k_z<k_F$ except $k_z=0$, which describes the Majorana-Fermi arc. Due to the singlet nature of the paired state, these surface states are two-fold spin degenerate, and the physical response of the chiral states such as anomalous spin and thermal Hall conductivities are indistinguishable from those of a complex chiral fermion. The Majorana-Fermi arc is shown as the dashed line in Fig.~\ref{fig:subfig8a} (also see Fig.~2(a)). In principle the Fermi arcs can be observed in both the ARPES and the Fourier transformed STM measurements. But, due to the issue of energy resolution of ARPES at low temperatures ($T<T_c \sim 1.5 K$), one may have to entirely rely on the STM measurements for the detection of the Majorana-Fermi arcs.

The chiral states also carry surface current along the $y$ direction. But, the total surface current is a sum of the contributions from the bound states and the scattered states \cite{Roy}. The local magnetization due to the surface current can be detected by SQUID microscopy and SQUID susceptometry \cite{Moler1,Moler2,Moler3}.

The anomalous spin and thermal Hall effects described by Eq.~(\ref{anomaly}), can be obtained by considering the response of these chiral surface states. When a gradient of the Zeeman coupling is applied, which acts as the spin electric field, the chiral surface states lead to an anomalous spin Hall current. Each chiral mode for a given value of $-k_F<k_z<k_F$ contributes $\hbar/(8 \pi)$ towards the spin Hall conductivity \cite{Fisher,Read}. After accounting for all the chiral modes we obtain the spin Hall conductivity at zero temperature to be
\begin{equation}\label{sh1}
\sigma^s_{xy,0}=\frac{\hbar}{8 \pi} \int_{-k_F}^{k_F} \frac{d k_z}{2\pi}=\frac{\hbar}{8 \pi} \times \left(\frac{k_F}{\pi}\right).
\end{equation} Similarly, each chiral Majorana mode at very low temperature contributes $\times \pi^2 k^2_B T/6h$ to the anomalous thermal Hall conductivity\cite{Fisher, Read}, and after accounting for all the modes (due to the allowed values of $k_z$ and the two-fold spin degeneracy) we end up with
\begin{equation}\label{th1}
\lim_{T \to 0} \; \kappa_{xy} = \frac{\pi^2 k^2_B T}{3h} \times \left(\frac{k_F}{\pi}\right).
\end{equation} Therefore, at low temperatures the spin and the thermal Hall conductivities satisfy a generalized Wiedemann-Franz law \cite{Vafek}
\begin{equation}\label{WF}
\lim_{T \to 0} \; \frac{\kappa_{xy}}{T \sigma_{xy}^{s}}=\frac{4 \pi^2 k^2_B}{3 \hbar^2}=L_s,
\end{equation}
where $L_s$ is a modified Lorenz number obtained by replacing electronic charge $e$ by $\hbar/2$. More detailed consideration of the anomalous response functions in the bulk will be discussed in Sec.~\ref{sec6} and Sec.~\ref{sec7}.

We conclude this subsection by discussing the effects of a uniform Zeeman coupling on the chiral surface states. The Zeeman splitting increases the gapless regions on the Fermi surface, and thereby decreases the number of chiral surface states. A uniform Zeeman coupling $\Delta_Z$ lifts the spin degeneracy and leads to the modified dispersion relations
\begin{equation}
E_s= \frac{\Delta_0}{k^2_F}|k_z| k_y + (-1)^s \Delta_Z,
\end{equation}
where $s$ represents the spin projections $\pm 1$. However, the wavefunctions for both projections are still described by Eq.~(\ref{wafn}). The spin-split Majorana-Fermi arcs are now described by the following hyperbolas
\begin{eqnarray}\label{MaF1}
k_{y,s}= (-1)^s \frac{\Delta_Z k^2_F}{\Delta_0 |k_z|},
\end{eqnarray} which are separated in the momentum space. By substituting $k_{y,s}$ in the normalizability criterion $k^2_y+k^2_z < k^2_F$, we obtain
\begin{eqnarray}
\frac{k_F}{\sqrt{2}}\sqrt{1-\sqrt{1-\frac{4\Delta^2_Z}{\Delta^2_0}}}<|k_z|<\frac{k_F}{\sqrt{2}}\sqrt{1+\sqrt{1-\frac{4\Delta^2_Z}{\Delta^2_0}}}, \nonumber \\
\end{eqnarray}
which shows the decrease in the total number of the chiral modes. This equation actually identifies the region on the Fermi surface, which is still gapped in the presence of the Zeeman coupling (recall that $k_z/k_F$ is the polar angle $\theta_{\mathbf{k}}$ on the Fermi surface). When $\Delta_Z > \Delta_0/2$, entire Fermi surface becomes gapless and the chiral modes are completely absent. The spin-split Majorana arcs for $\Delta_z/\Delta_0=0.4$ are shown in Fig.~\ref{fig:subfig8a}, as the solid lines. The evolution of the Majorana-arcs as a function of the Zeeman coupling can be tracked by the Fourier transformed STM measurements, which can provide valuable information regarding the topological nature of the pairing. Due to the decrease in the number of chiral modes, anomalous spin and thermal conductivities are reduced by the factor
\begin{eqnarray}
&&\frac{1}{\sqrt{2}}\sqrt{1+\sqrt{1-\frac{4\Delta^2_Z}{\Delta^2_0}}}-\frac{1}{\sqrt{2}}\sqrt{1-\sqrt{1-\frac{4\Delta^2_Z}{\Delta^2_0}}} \nonumber \\
&&\approx 1-\frac{\Delta_Z}{\Delta_0}, \: \: \: \mathrm{for} \: \: \: \: \Delta_z/\Delta_0 \ll \frac{1}{2}.
\end{eqnarray}

\section{Chiral Surface Andreev Bound States for $(k_x \pm ik_y)^2$ state}\label{sec5}
For the $(k_x \pm ik_y)^2$ state we expect the presence of chiral surface states on the $(1,0,0)$ and $(0,1,0)$ surfaces. In particular, we discuss the chiral surface states of $(k_x-ik_y)^2$ state on the $(1,0,0)$ surface. We consider a boundary in the x-direction, such that $x<0$ and $x>0$ regions are respectively occupied by the superconductor and the vacuum. The spinor wave-function now satisfies the differential equations
\begin{eqnarray}
&&\bigg[\left(-\partial^2_x+k^2_{y}+k^2_z-k^2_F\right)\tau_3 -\frac{\Delta }{\mu}\;\left(\partial^2_x +k^2_y \right)\tau_1 \nonumber \\ &&- \frac{2i\Delta }{\mu}k_y \partial_x\tau_2\bigg]\psi =2mE \psi,
\end{eqnarray}
and the boundary conditions $\psi(0)=\psi(x\to -\infty)=0$. We notice that $-\partial^2_x$ and $\partial_x$ respectively appear with the matrices $\left(\tau_3+\frac{\Delta_0}{\mu}\tau_1\right)$ and $\tau_2$. The calculation is  simplified by rewriting
\begin{eqnarray}
\left(-\partial^2_x+k^2_{y}+k^2_z-k^2_F\right)\tau_3 -\frac{\Delta }{\mu}\;\left(\partial^2_x +k^2_y \right)\tau_1 \nonumber \\
 =a \left(\tau_3+\frac{\Delta_0}{\mu}\tau_1\right)+b \left(\tau_1- \frac{\Delta_0}{\mu}\tau_3\right),
\end{eqnarray}
where
\begin{eqnarray}
&&a=-\partial^2_x+\frac{k^2_z+k^2_y(1-\Delta^2_0/\mu^2)-k^2_F}{1+\Delta^2_0/\mu^2}, \\
&&b=\frac{\Delta_0}{\mu}\frac{k^2_F-k^2_z-2k^2_y}{1+\Delta^2_0/\mu^2}.
\end{eqnarray}
Notice that $\mathcal{U}_3=\left(\tau_1- \frac{\Delta_0}{\mu}\tau_3\right)$ anticommutes with the Hamiltonian or the differential operator when $b=0$, and the zero energy states are thus protected by the spectral symmetry with respect to $\mathcal{U}_3$. The zero energy states are located at
\begin{equation}
k_y=\pm \frac{1}{2}\sqrt{k^2_F-k^2_z},
\end{equation}
and describe two spin degenerate Majorana-Fermi arcs. These arcs are shown by dashed lines in Fig.~\ref{fig:subfig8b} (also see Fig.~2(b)).

The index theorem mandates the existence of two independent zero energy solutions (without taking into account the spin degeneracy), which are eigenstates of $\mathcal{U}_3$ with eigenvalues $\mathrm{sgn}(k_y) \times \sqrt{1+\Delta^2_0/\mu^2}$. The differential equations are now solved by following the strategy of the previous subsection, and we obtain the following chiral dispersion relations (when $b \neq 0$) and the eigenfunctions
\begin{eqnarray}
&&E=-\mathrm{sgn}(k_y) \Delta_0 \sin \varphi \left(1-\frac{k^2_z}{k^2_F}-\frac{2k^2_y}{k^2_F}\right) \label{energy}\\
&&\psi_+(x)=A_3 \Theta(-x) \exp \left(|k_y| \; \cos \varphi \; x \right) \sin \bigg(\sin \varphi  x  \nonumber \\ && \times \sqrt{k^2_F-k^2_z-k^2_y}\bigg)\bigg[\Theta(k_y)\left(\begin{array}{c}
\cos \frac{\varphi}{2} \\
-\sin \frac{\varphi}{2}  \end{array} \right )+\Theta(-k_y) \nonumber \\
&& \: \: \: \: \: \: \: \: \: \: \: \: \: \: \: \: \: \: \: \: \: \: \times \left(\begin{array}{c}
\sin \frac{\varphi}{2} \\
\cos \frac{\varphi}{2}  \end{array} \right )\bigg]\label{eigenfunction},
\end{eqnarray}
where $\tan \varphi=\mu/\Delta_0$, and $A_3$ is the normalization constant. By setting $k_z=0$, and $\sin \varphi \sim 1$ in Eq.~(\ref{energy}) we recover the dispersion of the chiral edge modes of the corresponding two dimensional problem, which has been found in Ref.~\onlinecite{Golub}. If we linearize in the vicinity of the Fermi arcs, we obtain two chiral, spin-degenerate linearly dispersing Majorana modes \cite{Fisher,Golub}. At zero temperature the chiral surface states give rise to the spin Hall conductivity
\begin{equation}\label{sh2}
\sigma^s_{xy,0}=\frac{\hbar}{8 \pi} \int_{-k_F}^{k_F} \frac{d k_z}{2\pi}=\frac{\hbar}{8 \pi} \times 2 \times \left(\frac{k_F}{\pi}\right),
\end{equation}
which is two times larger than the $\sigma^s_{xy,0}$ for the $k_z(k_x \pm i k_y)$ state in Eq.~(\ref{sh1}). Similarly the low temperature thermal Hall conductivity
\begin{equation}\label{th1}
\lim_{T \to 0} \; \kappa_{xy} = \frac{\pi^2 k^2_B T}{3h} \times 2 \times \left(\frac{k_F}{\pi}\right)
\end{equation}
also turns out to be two times larger than that of $k_z(k_x \pm i k_y)$ state in Eq.~(\ref{th1}).

In the presence of a uniform Zeeman coupling, the dispersion relations are given by
\begin{eqnarray}
E_s=-\mathrm{sgn}(k_y) \Delta_0 \sin \varphi \left(1-\frac{k^2_z}{k^2_F}-\frac{2k^2_y}{k^2_F}\right)+(-1)^s \Delta_Z, \nonumber \\
\end{eqnarray}
from which we obtain the spin-split Majorana arcs. For practicality we can assume $\sin \varphi \sim 1$, and find four Majorana arcs described by
\begin{eqnarray}
k_{y}=\pm \frac{k_F}{\sqrt{2}}\left[1 \pm \frac{\Delta_Z}{\Delta_0}-\frac{k^2_z}{k^2_F}\right]^{\frac{1}{2}},
\end{eqnarray} when
\begin{eqnarray}
-k_F\sqrt{1-\frac{\Delta_Z}{\Delta_0}}<k_z< k_F\sqrt{1-\frac{\Delta_Z}{\Delta_0}},
\end{eqnarray} and $\Delta_z/\Delta_0 < 1$. These arcs are shown as solid lines in Fig.~\ref{fig:subfig8b}, for $\Delta_Z/\Delta_0=0.4$. Due to the decrease in the total number of the chiral modes, anomalous Hall conductivities are reduced by the factor
\begin{eqnarray}
\sqrt{1-\frac{\Delta_z}{\Delta_0}} \approx 1-\frac{\Delta_z}{2\Delta_0},
\end{eqnarray} for a small Zeeman coupling. Since $m=\pm 2$ pairing is more effective in gapping out Fermi surface in comparison to the $m=\pm1$ pairing, the chiral surface states for the $m=\pm 2$ state persist up to a stronger Zeeman coupling.
\begin{figure}[htb]
\includegraphics[width=8cm,height=6cm]{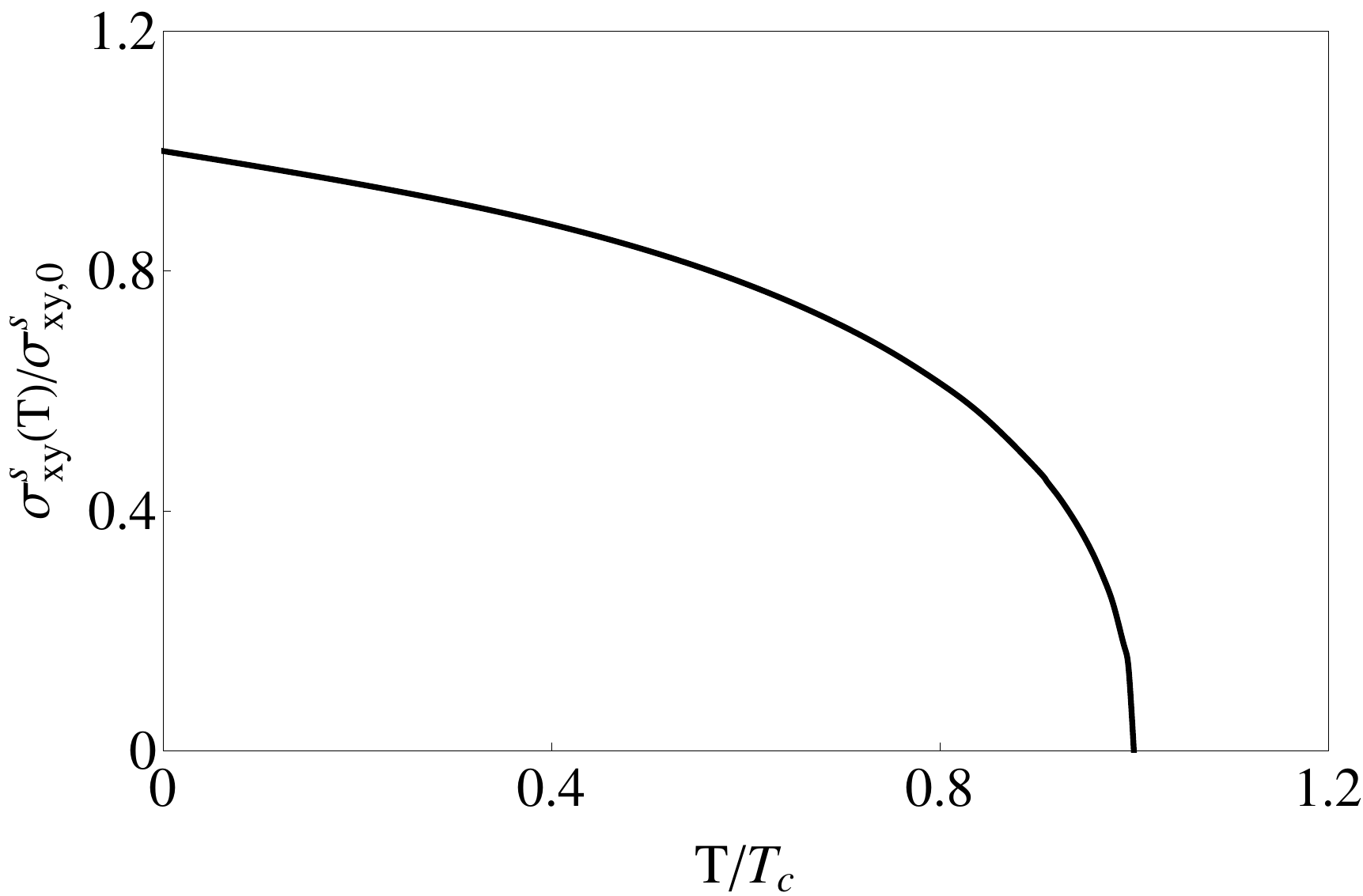}
\caption{(Color online)Temperature dependence of the anomalous spin Hall conductivity for $(k_x \pm i k_y)^2$ pairing, where $\sigma^s_{xy,0}= \frac{\hbar }{8 \pi} \times m \times \left(\frac{k_F}{\pi}\right)$, and $m=\pm 2$. A similar temperature dependence is also found for $k_z(k_x \pm i k_y)$ pairing, with $\sigma^s_{xy,0}= \frac{\hbar }{8 \pi} \times m \times \left(\frac{k_F}{\pi}\right)$, and $m=\pm 1$. }
\label{fig:6}
\end{figure}
\section{Anomalous Spin Hall conductivity}\label{sec6}
Due to the global spin rotational invariance of the singlet paired states, the spin is a conserved quantity and the spin conductivities can be computed as a linear response of the system in the presence of external spin gauge fields in the following way. We can work with either SU(2) or U(1) gauge fields. For simplicity, we choose the spin quantization axis along the $z$ direction, and work with the U(1) spin gauge fields, which are minimally coupled to the BCS quasiparticles, through the Peierls substitution $\mathbf{k} \to -i\nabla-\frac{1}{2} \mathbf{A}_s$ in Eq.~(\ref{hamiltonian1}) and Eq.~(\ref{hamiltonian2}). The Lagrangian density acquires the following form
\begin{eqnarray}
&&\mathcal{L}[\Psi,\Psi^\dagger, A_{\mu,s}] \nonumber \\
&&= \Psi^\dagger \bigg \{i \hbar \partial_t-\frac{\hbar A_{0,s}}{2}-\hat{h}\left[-i\nabla-\frac{\mathbf{A}_s}{2} \right]\bigg \}\Psi,
\end{eqnarray}
and the spin current vertices in the momentum space are given by
\begin{equation}
J_{a,s}=\frac{1}{2} \; \boldsymbol \tau \cdot \frac{\partial \mathbf{N}_{\mathbf{k}}}{\partial k_a }.
\end{equation}
It is important to note that the Zeeman coupling serves as the scalar potential of the spin gauge field. Therefore, a spatially varying Zeeman coupling gives rise to the spin electric field $\mathbf{E}_s=\nabla A_{0,s}$. The spin conductivities can be calculated either by using Kubo formulas or by using the following semiclassical equations
\begin{figure*}[htb]
\centering
\subfigure[]{
\includegraphics[width=8.6cm,height=7cm]{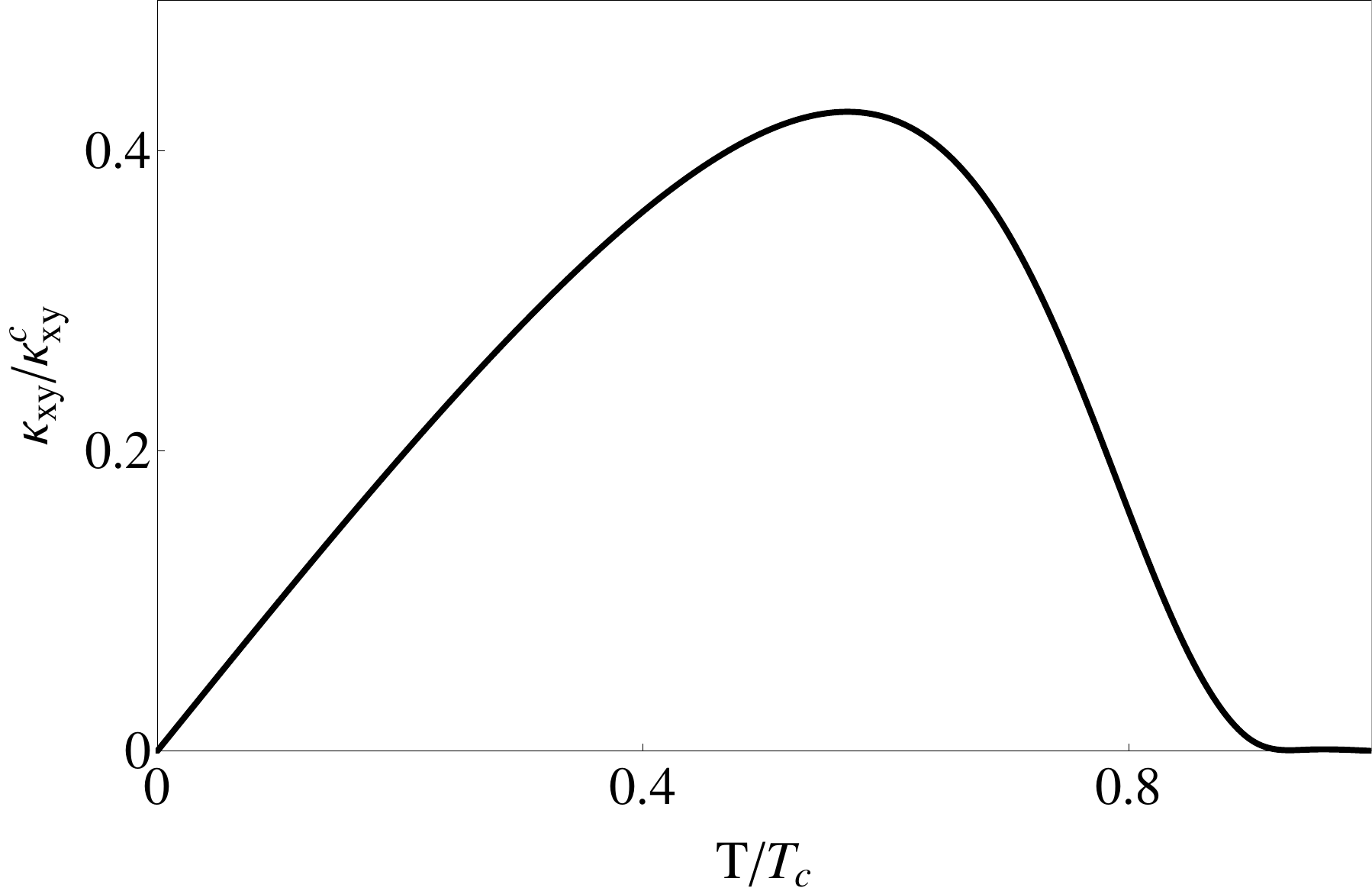}
\label{fig:subfig7a}
}
\subfigure[]{
\includegraphics[width=8.6cm,height=7cm]{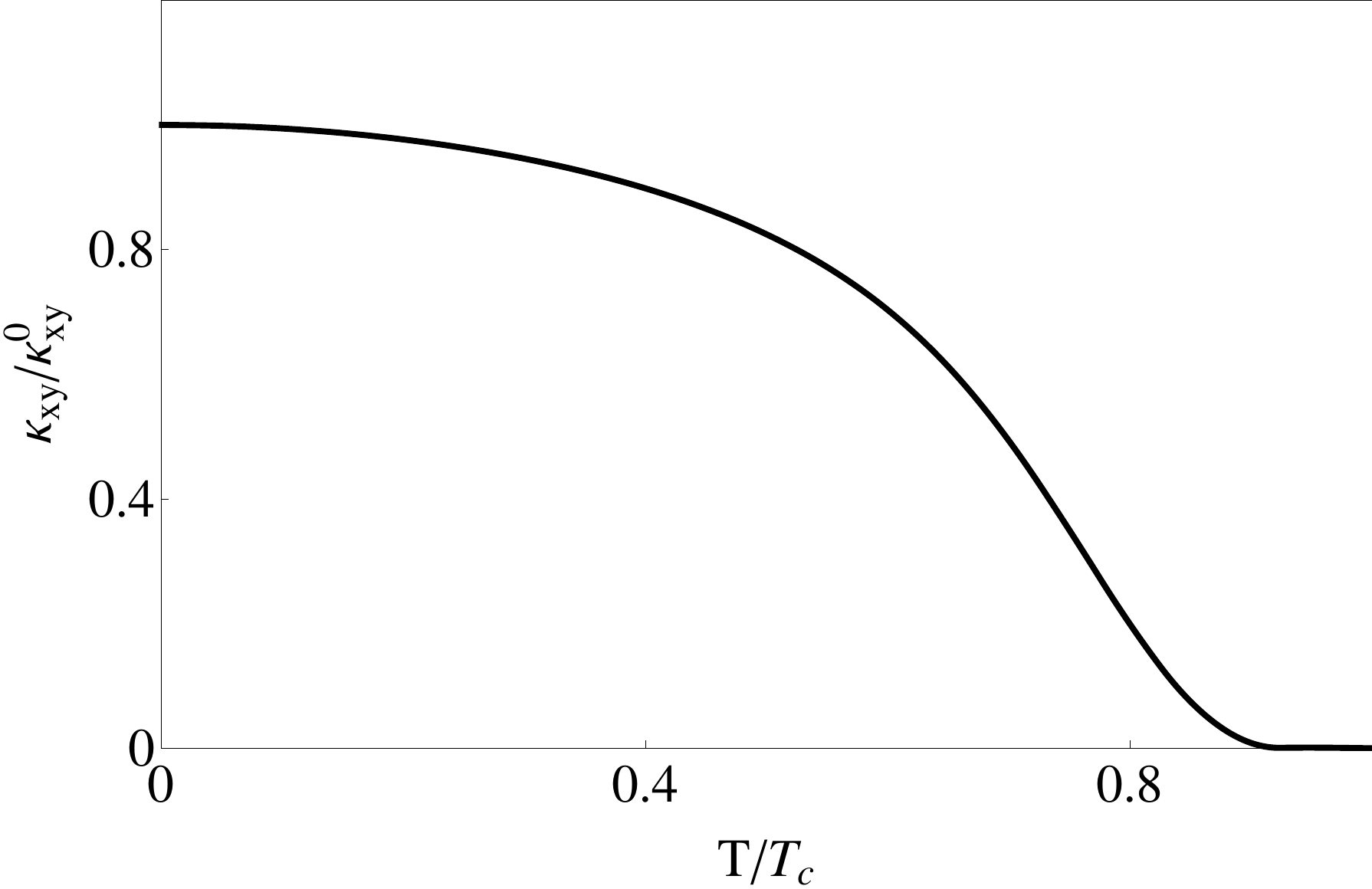}
\label{fig:subfig7b}
}
\label{fig:7}
\caption[]{(a) Temperature dependence of the anomalous thermal Hall conductivity $\kappa_{xy}$ for $(k_x \pm i k_y)^2$ pairing, where $\kappa^c_{xy}=\frac{\pi^2 k^2_B T_c}{3h} \times m \times \left(\frac{k_F}{\pi}\right)$, and $m=\pm 2$. $\kappa^c_{xy}$ roughly determines the maximum value of $\kappa_{xy}$ which occurs at a model dependent intermediate temperature $T \sim T_c/2$. (b) Temperature dependence of $\kappa_{xy}/\kappa^0_{xy}$ for $(k_x \pm i k_y)^2$ pairing, where $\kappa^0_{xy}=\frac{\pi^2 k^2_B T}{3h} \times m \times \left(\frac{k_F}{\pi}\right)$ is the low temperature limit of $\kappa_{xy}$, and $m=\pm 2$ . A similar temperature dependence of the $\kappa_{xy}$ is found for the $k_z(k_x \pm i k_y)$ pairing.  }
\end{figure*}
\begin{eqnarray}
\hbar \dot{\mathbf{r}}_{n}=\nabla_{\mathbf{k}} E_{n,\mathbf{k}}+ \hbar \dot{\mathbf{k}} \times \mathbf{\Omega}_{n,\mathbf{k}}, \\
\hbar \dot{\mathbf{k}}_{n}=\frac{\hbar}{2}\mathbf{E}_s+\frac{h}{2} \dot{\mathbf{r}}_{n} \times \mathbf{B}_s.
\end{eqnarray}Notice that the semiclassical equations are same as the ones used for describing the anomalous charge Hall effects in metals \cite{XiaoNiu,OngNagaosa}, with the replacements $e \to \hbar/2$, $\mathbf{E} \to \mathbf{E}_s$, and $\mathbf{B} \to \mathbf{B}_s$. In the absence of the ``spin magnetic field" $\mathbf{B}_s=\nabla \times \mathbf{A}_s$, the elimination of $\dot{\mathbf{k}}_{n}$ leads to
\begin{eqnarray}
\hbar \dot{\mathbf{r}}_{n}=\nabla_{\mathbf{k}} E_{n,\mathbf{k}}+ \frac{\hbar}{2} \mathbf{\Omega}_{n,\mathbf{k}}\times \mathbf{E}_s,
\end{eqnarray}
from which we immediately find the dc spin Hall conductivity
\begin{eqnarray}\label{sh3}
\sigma^s_{ab}=\left(\frac{\hbar}{2}\right)^2 \; \frac{1}{\hbar} \; \epsilon_{abc} \int \frac{d^3k}{(2\pi)^3} \; \Omega_{n,\mathbf{k},c} \; f(E_{n,\mathbf{k}}).
\end{eqnarray}
In the above formula we have implied summation over the repeated indices $n$, and $c$, and $f(\epsilon)=[1+\exp(\epsilon/k_B T)]^{-1}$
is the Fermi function. Since nonzero Berry flux passes only through the $ab$ plane, only $\sigma^s_{xy}$ is nonzero. At zero temperature and in the absence of a uniform Zeeman coupling, the dc spin Hall conductivity in the $xy$ plane is given by
\begin{eqnarray}\label{sh4}
&&\sigma^s_{xy,0}=\frac{\hbar}{8} \int \frac{d^3 k}{(2\pi)^3} \; \mathbf{n}_{\mathbf{k}} \cdot \bigg(\frac{\partial \mathbf{n}_{\mathbf{k}}}{\partial k_x} \times \frac{\partial \mathbf{n}_{\mathbf{k}}}{\partial k_y}\bigg)\nonumber \\
&&= \frac{\hbar m}{16 \pi^2}  \int dk_z \Theta(k^2_F-k^2_z)= \frac{\hbar }{8 \pi} \times m \times \left(\frac{k_F}{\pi}\right), \nonumber \\
\end{eqnarray}
where $\hbar/(8 \pi)$ is the unit of the quantum spin Hall conductivity. This formula is in agreement with the spin Hall conductivity obtained from the chiral surface states in Eq.~(\ref{sh1}) and Eq.~(\ref{sh2}), and establishes the bulk-boundary correspondence for anomalous spin transport. The universal result for the spin Hall conductivity in Eq.~(\ref{sh4}) can also be found by analyzing the chiral anomaly of the linearized theory of Weyl or double-Weyl fermions \cite{Zyuzin3,Grushin,Qi,Goswami2}.

The formula in Eq.~(\ref{sh3}) also allows us to study the dependence of $\sigma^s_{xy}$ on temperature and uniform Zeeman coupling. A constant Zeeman coupling acts as the chemical potential and causes a gradual decrease of $\sigma^s_{xy}$ from the universal value $\sigma^s_{xy,0}$ \cite{Goswami2}, which has also been noted in the previous two subsections. In the absence of a uniform Zeeman coupling, the temperature dependence of the spin Hall conductivity can be expressed as
\begin{eqnarray}
\sigma^s_{xy}=\sigma^s_{xy,0}F_s \left(\frac{\Delta_0(T)}{k_BT}, \frac{\hbar \Omega_D}{k_BT}\right)
\end{eqnarray} where $F_s \left(\frac{\Delta_0(T)}{k_BT}, \frac{\hbar \Omega_D}{k_BT}\right)$ is a dimensionless scaling function. For $T \to 0$ and $T \to T_c$, $F_s \to 1$ and $F_s \to 0$ respectively. In Fig.~\ref{fig:6}, we have plotted $\sigma^s_{xy}(T)/\sigma^s_{xy,0}$ for the $(k_x \pm i k_y)^2$ pairing, as a function of $T/T_c$. We have used the dimensionless BCS coupling constant $g_{BCS}=0.116$, which leads to $k_B T_C \approx 0.02 \hbar \Omega_D$. A similar temperature dependence of the spin Hall conductivity is also found for the $k_z(k_x \pm i k_y)$ pairing.

The spin Hall effect can lead to (i) an accumulation of spin density/ spin voltage in the transverse direction, and (ii) also an induced electric field due to the dipole current. When the superconductor is placed in an external magnetic field, due to the Meissner screening an inhomogeneous magnetic field occurs naturally inside the superconductor. This leads to a gradient of the Zeeman coupling, and an inhomogeneous spin current. In principle the resultant spin accumulation can be detected by using a SQUID device. But, the measurement of the resulting electric field can be more challenging due to the screening effects (Thomas-Fermi screening in a metal).

\section{Anomalous Thermal Hall Conductivity}\label{sec7}

In the previous sections we have already showed that the chiral surface states lead to an anomalous thermal Hall effect, even in the absence of any external magnetic field. Compared to the anomalous spin Hall effect, it is easier to experimentally detect the anomalous thermal Hall effect. For this reason we consider the bulk thermal Hall conductivity formula in detail. For the calculation of thermal Hall conductivity, one needs to carefully take into account the effects of the energy magnetization. This is crucial for the cancellation of the unphysical divergent contributions, arising from the naive application of the Kubo formula \cite{Vafek,Niu1,Nagaosa}. After a lengthy calculation the following simple formula
\begin{equation}\label{tHall}
\kappa_{ab}=-\frac{k^2_B T}{\hbar} \epsilon_{abc}\int \frac{d^3 k}{(2\pi)^3} \; \Omega_{n,\mathbf{k},c} \; \int_{E_{n,\mathbf{k}}}^{\infty} dE \; E^2 \; \frac{\partial f(E)}{\partial E},
\end{equation}
is obtained, where the summation over the repeated indices $n$ and $c$ have been implied. By completing the integral over $E$, and also noticing that only $\kappa_{xy} \neq 0$, we arrive at
\begin{eqnarray}
&&\kappa_{xy}=-\frac{\pi^2 k^2_B T}{3\hbar}\int \frac{d^3 k}{(2\pi)^3} \; \Omega_{n,\mathbf{k},z} \; \bigg[1 + \frac{3}{\pi^2} \bigg \{\frac{E^2_{n,\mathbf{k}}}{k^2_B T^2}\times \nonumber \\ &&  f(E_{n,\mathbf{k}})-\log^2[1-f(E_{n,\mathbf{k}})] -2 \mathrm{Li}_2[1-f(E_{n,\mathbf{k}})]\bigg \} \bigg]. \nonumber \\
\end{eqnarray}
In the low temperature limit, we find $\kappa^0_{xy}= \frac{\pi^2 k^2_B T}{3h} \times m \times \left(\frac{k_F}{\pi}\right)$. In the absence of a uniform Zeeman coupling, the temperature dependence of $\kappa_{xy}$ over the entire range $0<T<T_C$ can be captured as
\begin{equation}
\kappa_{xy}= \kappa^0_{xy} \; F_\kappa \left(\frac{\Delta_0(T)}{k_BT}, \frac{\hbar \Omega_D}{k_BT}\right),
\end{equation}
where $F_\kappa$ is a scaling function. The $\kappa_{xy}$ vanishes linearly as $T \to 0$, and also vanishes as $T \to T_c$. The maximum value of $\kappa_{xy}$ is comparable to
\begin{equation}
\kappa^c_{xy}=\frac{\pi^2 k^2_B T_c}{3h} \times m \times \left(\frac{k_F}{\pi}\right),
\end{equation}
which is attained at a model dependent intermediate temperature $T \sim T_c/2$. We have plotted $\kappa_{xy}/\kappa^c_{xy}$ as a function of $T/T_c$ for $m= \pm 2$ pairing in Fig.~\ref{fig:subfig7a}. We have again used a dimensionless BCS coupling constant $g_{BCS}=0.116$. We have also plotted the ratio $\kappa_{xy}/\kappa^0_{xy}$, as a function of temperature in Fig.~\ref{fig:subfig7b}, which saturates to unity as $T \to 0$.
\begin{figure*}[htb]
\centering
\subfigure[]{
\includegraphics[width=8.6cm,height=7cm]{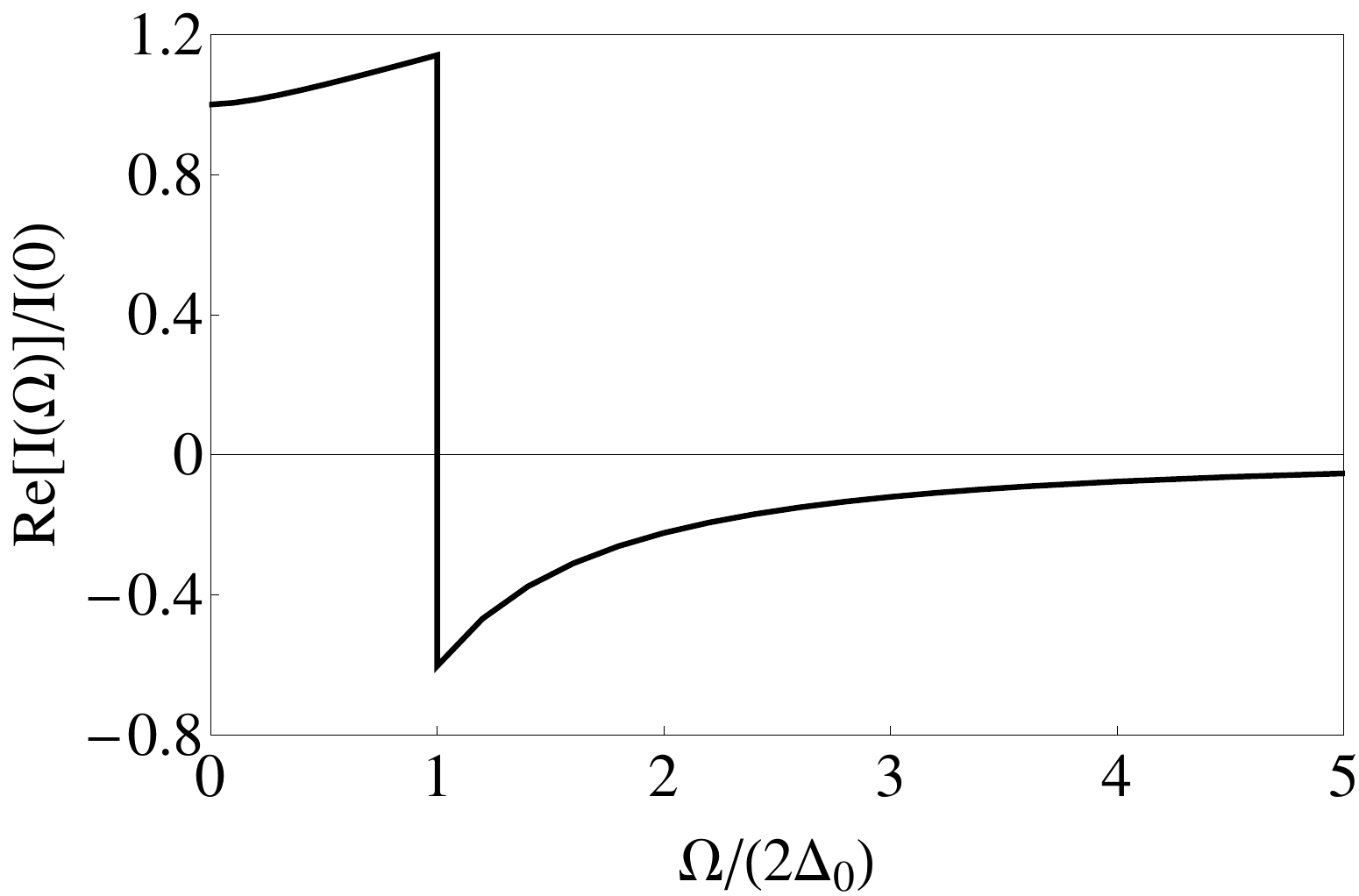}
\label{fig:subfig7a}
}
\subfigure[]{
\includegraphics[width=8.6cm,height=7cm]{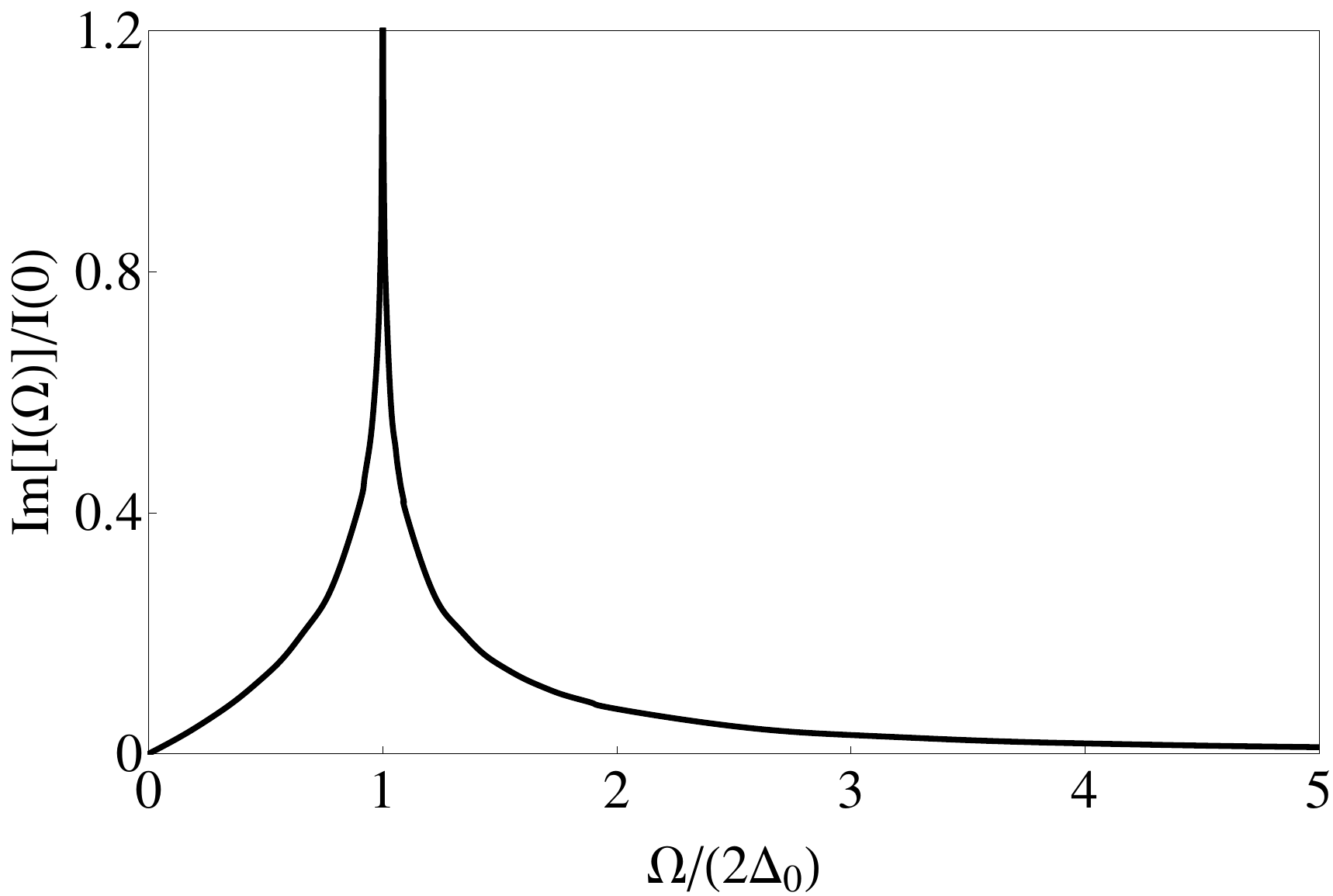}
\label{fig:subfig7b}
}
\label{fig:7}
\caption[]{(a) The real part of the antisymmetric density-current correlation function for $(k_x \pm ik_y)^2$ pairing, as a function of external frequency $\Omega$, which attains its maximum value at $\Omega=2\Delta_0$. At the same frequency the real part jumps discontinuously from positive to negative values. (b) The imaginary part of the antisymmetric density-current correlation function for $(k_x \pm ik_y)^2$ pairing, which diverges at $\Omega=2\Delta_0$. Due to the presence of gapless excitations with linear density of states, the imaginary part behaves as $\propto \Omega$ at low frequencies $\Omega \ll 2\Delta_0$. Here $I(0)$ corresponds to the real part at zero frequency. A similar behavior for the antisymmetric current-density correlation function is also found for the $k_z(k_x \pm i k_y)$ pairing.}
\end{figure*}

Now we consider the feasibility of the experimental detection of the anomalous $\kappa_{xy}$ for URu$_2$Si$_2$, which is a low $T_c$ superconductor. For this, we provide an estimate of the $\kappa^c_{xy}$. For real material we also need to account for multiple Fermi pockets, and possible disorder contributions. From the charge Hall conductivity measurements in the normal state (hidden ordered metallic phase HO in Fig.~\ref{fig:1}), an estimation of $k_F \sim 1.8 \times 10^9 m^{-1}$ has been obtained for the hole pocket (presumably around $Z$ point). Using this value of $k_F$ and $T_c \sim 1.5 K$, we obtain
\begin{equation}
\kappa^c_{xy} \sim m \times 8 \times 10^{-4} W K^{-1} m^{-1} \sim 10^{-3} W K^{-1} m^{-1} .
\end{equation}
This magnitude is comparable to the values of $\kappa_{xy}$ found in Ba$_{1-x}$K$_x$Fe$_2$As$_2$, in the presence of an external magnetic field $B \sim 2 T$ around $T \sim 35 K$ in Ref.~\onlinecite{Ong1} ($T_c$ for this material is $\sim 37 K$). Therefore, the large intrinsic anomalous thermal Hall conductivity of the possible chiral states are measurable with current experimental setup. The estimated value can be further increased due to the presence of additional Fermi pockets, and skew scattering and side jump contributions due to disorder.

\section{Summary and future directions}\label{sec11}
We have discussed that the current thermodynamic and thermal conductivity measurements in the superconducting state of URu$_2$Si$_2$ suggest the presence of exotic nodal quasiparticles with linear density of states\cite{Hasselbach,Kohori,Kasahara1,Kasahara2,Yano}. The magnetization \cite{GangLi} and the polar Kerr effect \cite{Kapitulnik} measurements provide the evidence for a time reversal symmetry breaking, chiral paired state. In addition the neutron scattering resonance observed at $\mathbf{Q}=(0,0,2\pi/c)$ reflects the existence of a momentum dependent pairing amplitude $\Delta_{\mathbf{k}}$, which changes sign under $\mathbf{k} \to \mathbf{k} +\mathbf{Q}$. We have showed that the nodal quasiparticles of both $k_z(k_x \pm i k_y)$ (see Eq.~(\ref{pairing1})) and $(k_x \pm i k_y)^2$ (see Eq.~(\ref{pairing2})) pairings can explain the current experimental results. However, these two states are topologically distinguished by their orbital angular momentum projections along the $c$ axis, which are respectively given by $m=\pm1$ and $m=\pm 2$. Due to the uncertainty over the nature of the hidden ordered normal state and its fermiology, it is difficult to choose between these two paired states. We have suggested that the bulk topological invariant $m$ can be determined by performing corner Josephson junction measurements.

We have showed that the angular momentum projections along the $c$-axis govern the Berry flux through the $ab$ plane, and the nature of the point nodes. The point nodes  of the $m=\pm 1$ and the $m=\pm 2$ pairings on the $c-axis$ respectively act as the unit and the double (anti)monopoles of the Berry curvature (see Fig.~\ref{fig:4} and Fig.~\ref{fig:5}), which in turn realize the Weyl (see Eq.~(\ref{Weyl})) and the double Weyl (see Eq.~(\ref{doubleWeyl})) fermions. The Weyl fermions possess linear dispersion along all three directions and its density of states vanishes quadratically with energy. In contrast, the double Weyl fermions have linear and quadratic dispersions along the $c$ axis and in the $ab$ plane respectively, and leads to a density of states, which vanishes linearly with energy (see Eq.~(\ref{DOSdoubleWeyl})). The linear density of states for $m=\pm 1$ pairing can arise due to the presence of line nodes on the Fermi surfaces around the $\Gamma$ or the $Z$ points. We have also demonstrated that a line node acts as the vortex loop in the momentum space and is protected by a topological invariant (see Eq.~(\ref{tinvring})), which is not tied with the angular momentum projection and is distinct from the topological invariant of the point nodes. The presence of nontrivial topological invariants for the nodal quasiparticles lead to protected zero energy surface states, and the bulk topological invariant can also be determined by probing the surface states through ARPES and Fourier transformed STM measurements. For this reason, we have discussed the possible surface states of both pairings in detail. 

The line node in the $ab$ plane gives rise to dispersionless, spin degenerate bound states on the $(0,0,1)$ surface, which produce an image of the equatorial cross-section of the Fermi surface (see Fig.~2(a)). In contrast the Berry curvature of the chiral paired state (hence the point nodes) lead to the chirally dispersing surface Andreev bound states on the $ca$ and the $cb$ plane. Their energies vanish along the Majorana-Fermi arcs, which are bounded by the images of the point nodes. The number of spin-degenerate Majorana-Fermi arcs equals $m$ and their profiles are showed by dashed lines in Fig.~\ref{fig:subfig8a} and Fig.~\ref{fig:subfig8b}. We have also discussed the effects of Zeeman coupling on the bound states. The dispersionless, zero energy bound states are immediately gapped out by the Zeeman splitting. The evolution of the Majorana-Fermi arcs due to the Zeeman splitting is showed by the solid lines in Fig.~\ref{fig:subfig8a} and Fig.~\ref{fig:subfig8b}. Therefore, a through analysis of these Majorana-Fermi arcs by Fourier resolved STM measurements can clearly establish the precise nature of the topological pairing. In addition the chiral surface states carry surface current, which can be probed by SQUID microscopy. 

The Berry curvature in bulk leads to anomalous spin and thermal Hall conductivities (see Eq.~(\ref{sh3}) and Eq.~(\ref{tHall})). In the low temperature limit the bulk results for these anomalous Hall conductivities precisely match with the results obtained from the chiral surface states, which demonstrates the bulk-boundary correspondence. The spin and the thermal Hall conductivities at low temperatures satisfy a generalized Wiedemann-Franz law (see Eq.~(\ref{WF})). The spin Hall effect arises in the presence of a gradient of the Zeeman coupling, and its direct measurement can be quite challenging. Perhaps the anomalous thermal Hall conductivity (in the absence of an external magnetic field) can be measured with relative ease. We have showed that the maximum value of thermal Hall conductivity $ \sim 10^{-3} WK^{-1}m^{-1}$ is attained around $T \sim T_c/2$, which is indeed measurable with current experimental setup.

For simplicity, we have worked with a Galilean invariant dispersion relation in the normal state. This is useful in pointing out the salient topological features of the paired state. However, it is inadequate for obtaining a correct description of the electrodynamic and the thermoelectric response. Within the Galilean invariant description, the momentum independent current-current correlation functions vanish and the difference between a chiral and a nonchiral pairings arise only through an antisymmetric current-density correlation function \cite{Roy,Golub,Goryo,GoryoIshikawa,Lutchyn,Kallin}. This term can give rise to various magnetoelectric effects \cite{GoryoIshikawa,Lutchyn}. After integrating out the phase degree of freedom, this anomalous current-density correlation function also leads to a momentum and frequency dependent anomalous charge Hall conductivity. But, the ac Hall conductivity (obtained by setting the external momentum zero) vanishes \cite{Golub,Goryo,Lutchyn,Kallin}. The estimate for polar Kerr angle obtained from the momentum and the frequency dependent Hall conductivity is very small due to the suppression by a factor of $(v/c)^2$ (see Ref.~\onlinecite{Lutchyn,Kallin}). Nevertheless, we have calculated the frequency dependent anomalous current-density correlation functions for the $m=\pm1$ and the $m=\pm2$ pairings, within the Galilean invariant formalism. The frequency dependence of the real and the imaginary parts of this correlation function for $m=\pm 2$ pairing are respectively shown in Fig.~\ref{fig:subfig7a} and Fig.~\ref{fig:subfig7b}, where $I(0)$ corresponds to the real part at zero frequency. In contrast to the fully gapped two dimensional chiral pairings discussed in Ref.~\onlinecite{Golub,Lutchyn,Kallin}, the imaginary part is finite for small frequencies such that $\Omega < 2 \Delta_0$. At very low frequency, due to the linear density of states of the nodal quasiparticles the imaginary part of the correlation function is $\propto \Omega$. For a proper estimation of the polar Kerr angle, it is important to consider the effects of impurities \cite{Goryo1,Lutchyn1}, and the multiband nature \cite{Kallin1} of the normal state. Both effects can give rise to a finite ac charge Hall conductivity, which can significantly enhance the polar Kerr angle. A detailed consideration of the electrodynamic properties by incorporating the multi-band nature of the hidden ordered normal state will be provided in a future publication. 

The Berry curvature can also lead to additional anomalous transport coefficients, such as spin and charge Nernst effects. A finite spin Nernst conductivity (and its inverse effect through Onsager's reciprocity) can be found (within the Galiliean invariant description) in the presence of a uniform Zeeman coupling by employing the semi-classical formalism of Ref.~\onlinecite{Niu2,Tewari}. This suggests that a thermal gradient can be obtained by applying a gradient of the Zeeman coupling and  vice versa, in the mixed state. By breaking the Galiliean invariance, a quasiparticle contribution to the anomalous charge-Nernst effect is also found. More details of these cross-correlated response will be discussed elsewhere. In conclusion, we have identified URu$_2$Si$_2$ as a promising material for exploring gapless topological superconductivity.

\section{acknowledgements}  P. G. is supported at the National High Magnetic Field Laboratory by NSF Cooperative Agreement No.DMR-0654118, the State of Florida, and the U. S. Department of Energy. L. B. is supported by DOE-BES through award DE-SC0002613.

\end{document}